\documentclass[twoside]{article}
\usepackage{color}
\usepackage{amssymb}
\usepackage{dsfont}
\usepackage{amssymb,amsmath,ulem,cancel}
\numberwithin{equation}{section}

\newtheorem{theorem}{Theorem}[section]
\newtheorem{lem}{Lemma}[section]
\newtheorem{pro}{Proposition}[section]
\newtheorem{cor}{Corollary}[section]
\newtheorem{rem}{Remark}[section]
\newtheorem{rems}{Remarks}[section]
\newtheorem{ex}{Example}[section]
\newtheorem{defi}{Definition}[section]
\newtheorem{hyp}{Assumption}[section]
\newtheorem{con}{Conjecture}[section]

\newcommand{\ssc}{\subsection}
\newcommand{\sssc}{\subsubsection}

\newcommand{\bt}{\begin{theorem}}
\newcommand{\et}{\end{theorem}}
\newcommand{\bl}{\begin{lem}}
\newcommand{\el}{\end{lem}}
\newcommand{\bp}{\begin{pro}}
\newcommand{\ep}{\end{pro}}
\newcommand{\bcor}{\begin{cor}}
\newcommand{\ecor}{\end{cor}}
\newcommand{\bcon }{\begin{con} \rm }
\newcommand{\econ }{\end{con}}
\newcommand{\lab }{\label }
\newcommand{\bd}{\begin{defi} \rm }
\newcommand{\ed}{\end{defi}}
\newcommand{\brem }{\begin{rem} \rm }
\newcommand{\erem }{\end{rem}}
\newcommand{\brems }{\begin{rems} \rm }
\newcommand{\erems }{\end{rems}}
\newcommand{\bhyp }{\begin{hyp} \rm }
\newcommand{\ehyp }{\end{hyp}}
\newcommand{\bex}{\begin{ex} \rm }
\newcommand{\eex}{\end{ex}}
\newcommand{\be}{\begin{equation}}
\newcommand{\ee}{\end{equation}}
\newcommand{\bde}{\begin{displaymath}}
\newcommand{\ede}{\end{displaymath}}
\newcommand{\beq}{\begin{eqnarray*}}
\newcommand{\eeq}{\end{eqnarray*}}
\newcommand{\beqa}{\begin{eqnarray}}
\newcommand{\eeqa}{\end{eqnarray}}
\newcommand{\bea}{\begin{align*}}
\newcommand{\eea}{\end{align*}}

\def\proof{\noindent {\it Proof. $\, $}}
\def\endproof{\hfill $\Box$ \vskip 5 pt}

\def\I{\mathds{1}}
\def\wh{\widehat}
\def\wt{\widetilde}
\def\phi{\varphi }

\newcommand{\pA}{A}
\newcommand{\pC}{C}
\newcommand{\pCC}{H}
\newcommand{\pD}{D}
\newcommand{\pAC}{A^{\pC}}
\newcommand{\pFC}{F^{\pC}}
\newcommand{\pDC}{D^{\pC}}
\newcommand{\pl}{l}

\newcommand{\Scum}{S^{\rm cum}}

\def\t1{\tau_{(1)}}

\def\gg{{\mathbb G}}

\def\F{{\cal F}}

\def\G{{\cal G}}

\def\P{\mathbb P}
\def\PS{{\mathbb P}^*}

\def\PT{\wt {\mathbb P}}

\def\E{{\mathbb E}}

\def\EPS{{\mathbb E}_{\mathbb  P^*}}
\def\EPG{{\mathbb E}_{\mathbb  P^{\gamma }}}

\def\FVA{{\rm FVA}}
\def\CVA{{\rm CVA}}
\def\DVA{{\rm DVA}}

\def\C-FVA{{\rm C-FVA}}

\def\sc{\textcolor{red}}

\title{{\Large \bf VALUATION AND HEDGING OF OTC 
CONTRACTS \\ \vskip 5 pt WITH 
FUNDING COSTS,
COLLATERALIZATION \\ AND COUNTERPARTY CREDIT RISK: PART 1}\vskip 60 pt}

\author{Tomasz R. Bielecki\footnote{The research of Tomasz R. Bielecki was supported under NSF grant DMS-1211256.}
\\ Department of Applied Mathematics \\
 Illinois Institute of Technology \\
 Chicago, IL 60616, USA \\ \\
Marek Rutkowski\footnote{The research of Marek Rutkowski was supported under Australian Research Council's Discovery Projects funding scheme (DP120100895).}
\\ School of Mathematics and Statistics \\ University of Sydney
\\ Sydney, NSW 2006, Australia}

\date{\vskip 35 pt Version of June 21, 2013. \vskip 10 pt}

\begin{document}
\maketitle
\vskip 40 pt
\begin{abstract}
The research presented in this work is motivated by recent papers by Brigo et al. \cite{BCP12,BCPP11,BPP11},
Cr\'epey \cite{SC1,SC2}, Burgard and Kjaer \cite{BK}, Fujii and Takahashi \cite{FT}, Piterbarg \cite{PV10}
and Pallavicini et al. \cite{PPB12}. Our goal
is to provide a sound theoretical underpinning for some results presented in these papers
by developing a unified martingale framework for the non-linear approach to hedging and pricing of OTC financial contracts.
The impact that various funding bases and margin covenants exert on the
values and hedging strategies for OTC contracts is examined.
The relationships between our research and papers by other authors, with an exception of Piterbarg \cite{PV10} and
Pallavicini et al. \cite{PPB12}, are not discussed in this Part 1 of our research. More detailed studies of these
relationships and modeling issues are examined in the follow-up Part 2.
\end{abstract}

\newpage
\tableofcontents
\newpage

\section{Introduction}


The aim of our research is to built a framework for valuation and hedging of OTC contracts between two (or, in perspective, more than two) defaultable counterparties in the presence of funding costs, collateralization and netting.
In the first part of the paper, the goal is to derive general results for wealth dynamics of trading strategies
in a market model where funding costs are specific to each party. In addition, we cover the
issues of benefits and losses at default, netting of positions, and various covenants regarding
the margin account in the case of collateralized contracts. We thus hope to develop a fairly general framework,
which can be applied to a wide range of models and problems arising in practice.
In the second part of the paper, we will apply our general results to problems arising in the market practice, specifically, to valuation of contracts under different funding costs for the defaultable counterparties.
Following the existing terminological convention, the valuation problem is abbreviated henceforth as FCVA (funding and credit valuation adjustment), although it is reasonable to argue that we simply deal here with the fair valuation of a contract under specific, sometimes quite complicated, trading rules. The commonly used term `adjustment' refers to a comparison of solutions to the valuation problem obtained using at least two different set-ups, and this is not necessarily our goal. We would thus like to stress that our approach should be contrasted with the heuristic `additive adjustments approach', which hinges on the additive price decomposition
\be\label{cacy}
 \wh \pi    = \pi  + \CVA + \DVA + \FVA + \textrm{additional adjustments (if needed)}
\ee
where $\pi $ is the fair value of the uncollateralized contract between non-defaultable counterparties and $\wh \pi $ is the `value' for the investor of the contract between two defaultable parties with idiosyncratic funding costs, collateral, and other relevant costs and/or risks. In most existing papers, the authors attempted to obtain explicit representation for the price decomposition \eqref{cacy} using three tools: \hfill \break (a) a thorough
 analysis of the contract's future cash flows, \hfill \break (b) some (rather arbitrary) choice of discounting of futures cash flows, and \hfill \break (c) the postulate that the risk-neutral valuation can be applied, so that the price is computed as the (conditional) expectation of discounted cash flows.

 It is still uncommon in this area to directly refer to hedging arguments, although this
 technique is used in some recent works.
A simple decomposition of a contract into a sequence of cash flows is justified when one deals with a contract in which cash flows themselves are independent of hedging strategies or, equivalently, the yet unknown
value process of the contract. This postulate is manifestly wrong when one deals with a collateralized
contract, in which the collateral amount is given in terms of the `fair' (or market) value of the contract.
By the same token, a particular form of discounting was usually adopted as a plausible postulate,
rather than derived as a strict result starting from some fundamental arguments. Obviously, any ad hoc choice of discounting
is questionable. Therefore, the practical approach summarized briefly above is manifestly flawed on numerous counts, though it may still sometimes yield a correct answer, provided that the problem is simple enough, so that a sensible solution is readily available anyway to a skilled quant. First, discounting using risk-free rate is reasonable only when the risk-free bond is traded.
Otherwise, when multiple yield curves are present, the choice of a discount factor is not arbitrary
and thus this issue needs to be carefully addressed.
Second, risk-neutral valuation is justified only when the wealth process of a hedging strategy is a martingale
under some probability measure after suitable discounting, so that the analysis of the drift term in the wealth dynamics is another crucial step and it can only be done by considering first trading strategies for both
 counterparties. In addition, it is well known that in the case of a non-linear pricing rule (or, equivalently, a solution to
  a non-linear backward stochastic differential equation (BSDE)), the discounted wealth process is not a martingale, so the classic approach to arbitrage pricing does not apply.
Third, although the choice of a num\'eraire asset is in principle arbitrary, the choice of the discount
factor should be consistent with the actual dynamics of the wealth under the statistical probability measure. Even when one insists on the choice of a conventional `risk-free rate'
as a discount factor, the problem of finding the wealth dynamics under the corresponding martingale measure remains a crucial issue that need to be analyzed in detail.
As already mentioned, the wealth dynamics will usually depend on the choice of a hedging strategy
and thus instead of computing the conditional expectation, one needs to
solve a non-linear BSDE.

\newpage

\section{Trading Strategies and Wealth Dynamics}\label{mm}

A finite trading horizon $T$ for our model of the financial market is fixed throughout the paper.
All processes introduced in what follows are implicitly assumed to be given on the underlying probability space $(\Omega, \G, \gg , \P)$ where the filtration $\gg = (\G_t)_{t \in [0,T]}$ models the flow of information available to all traders. We denote by $S^i$ the {\it ex-dividend price} (or simply the {\it price}) of the $i$th risky security with the cumulative dividend stream after time 0 represented by the process $\pA^i$.
Let $B^i$ stand for the corresponding {\it funding account} representing either unsecured or secured funding for the $i$th asset. A more detailed financial interpretation of these accounts will be discussed later.
We also introduce the {\it cash account} $B^0$, which is used for unsecured lending or borrowing of cash.
In the case when the borrowing and lending cash rates are different, we will use symbols $B^{0,+}$ and $B^{0,-}$
to denote the processes modeling unsecured lending and borrowing cash accounts, respectively.
A similar convention will be applied to processes $B^{i,+}$ and $B^{i,-}$.
For any random variable $\chi $, the equality $ \chi = \chi^+ - \chi^-$
is the usual decomposition of a random variable $\chi$ into its positive and negative parts.
Note, however, that this convention does not apply to double indices, such as $^{0,+}$ or $^{0,-}$.

\bhyp
We assume that: \hfill \break
(i) $S^i$ for $i=1,2,\dots , d$ are c\`adl\`ag semimartingales, \hfill \break
(ii) $\pA^i$ for $i=1,2,\dots , d$ are c\`adl\`ag processes of finite variation with $\pA^i_{0}=0$, \hfill \break
(iii) $B^j$ for $j=0,1,\dots , d$ are strictly positive and continuous processes of finite variation with $B^j_0=1$.
\ehyp

The {\it cumulative-dividend price} $S^{i,\textrm{cld}}$ is given as
\be \lab{pri1}
S^{i,\textrm{cld}}_t := S^i_t+ B^{i}_t\int_{(0,t]} (B^i_u)^{-1} \, d\pA^i_u ,\quad t\in [0,T],
\ee
and thus the {\it discounted cumulative-dividend price}  $\wh S^{i,\textrm{cld}}:=  (B^i)^{-1} S^{i,\textrm{cld}}$ satisfies
\be \lab{pri2}
\wh S^{i,\textrm{cld}}_t = \wh S^i_t+\int_{(0,t]} (B^i_u)^{-1} \, d\pA^i_u ,\quad t\in [0,T] ,
\ee
where we denote $\wh S^i :=  (B^i)^{-1} S^i$. If the $i$th traded asset does not pay any dividend up to time $T$ then the equality $ S^{i,\textrm{cld}}_t = S^{i}_t$ holds for every $t \in [0,T]$.
Note that the processes $S^{i,\textrm{cld}}$, and thus also the processes $\wh S^{i,\textrm{cld}}$, are c\`adl\`ag.


 Note that formula (\ref{pri1}) hinges on an implicit assumption that positive (resp. negative) dividends from the $i$th asset are invested in (resp. funded from) the $i$th funding account $B^i$. Since the main valuation and hedging results for derivative securities obtained in this section are represented in terms of primitive processes $S^i,B^i$ and $\pA^i$, rather than $S^{i,\textrm{cld}}$, our choice of a particular convention regarding reinvestment of dividends associated with the asset $S^i$ is immaterial.

\ssc{Trading Strategies and Funding Costs} \lab{secfun}

We are in a position to introduce trading strategies based on a family of traded assets introduced above.
 In this preliminary section, we mainly focus on definitions and notation used in what follows.

\bhyp
 We assume that $\xi^i$ for $i=1,2,\ldots,d$ (resp. $\psi^j$ for $j=0,1,\dots ,d$) are arbitrary $\gg$-predictable
(resp. $\gg$-adapted) processes such that the stochastic integrals used in what follows are well defined.
\ehyp

 In Sections \ref{secfun} and \ref{secelem}, we consider a dynamic portfolio $\phi =
 (\xi,\psi) = (\xi^1,\dots ,\xi^d, \psi^0,\dots ,\psi^d)$  composed of
risky securities $S^i,\, i=1,2,\ldots,d$, the cash account $B^0$ used for unsecured lending/borrowing,  and
funding accounts $B^i,\, i=1,2,\ldots,d,$ used for (unsecured or secured) funding of the $i$th asset.
In addition, we introduce a c\`adl\`ag process $\pA$ of finite variation, with $\pA_{0} = 0$, which is aimed to
represent the {\it external cash flows}, that is, the cash flows associated with some OTC contract.

\newpage

\brem
In the financial interpretation, the process $\pA$ is aimed to model all contractual cash flows either paid
out from the wealth or added to the wealth, as seen from the perspective of the {\it hedger} (the other
party is referred to as the {\it counterparty}). The name {\it contractual cash flows}
is used to emphasize that the process $\pA$ will typically model all cash flows directly generated by a security to be
replicated by means of a trading strategy $\phi.$
\erem

The wealth of a trading strategy depends on both $\phi $ and $\pA$, as is apparent from the next definition.
It is important to stress that hedging strategy $\phi $ and external cash flows $\pA$ cannot be separated, in general, since the wealth will depend in a non-linear way on both $\phi $ in $\pA$, in general.

\bd  \lab{ts1}
We say that a {\it trading strategy} $(\phi , \pA )$ with cash flows $\pA$ is  {\it self-financing}
whenever the wealth process $V(\phi , \pA )$, which is given by the formula
\be  \lab{rtf1}
V_t (\phi , \pA ) :=  \sum_{i=1}^d \xi^i_tS^i_t + \sum_{j=0}^d \psi^j_t B^j_t ,
\ee
satisfies, for every $t \in [0,T]$,
\be  \lab{portf2}
V_t (\phi , \pA ) = V_0(\phi , \pA )+ \sum_{i=1}^d \int_{(0,t]} \xi^i_u \, d(S^i_u + \pA^i_u )
+  \sum_{j=0}^d \int_{(0,t]} \psi^j_u \, dB^j_u + \pA_t
\ee
where $V_0(\phi , \pA )$ is an arbitrary real number.
\ed

\brem
Observe that (\ref{portf2}) yields the following wealth decomposition
\be \lab{c2a}
V_t (\phi , \pA ) = V_0(\phi , \pA )+ G_t(\phi , \pA ) + F_t(\phi  , \pA ) + \pA_t
\ee
where
\be \lab{gg66}
G_t (\phi , \pA ):=\sum_{i=1}^d  \int _{(0,t]} \xi^i_u \, (dS^i_u + d\pA^i_u)
\ee
represents the gains (or losses) associated with holding long/short positions in risky assets $S^1, \dots , S^d$ and
\be
F_t (\phi , \pA ) := \sum_{j=0}^d \int_{(0,t]} \psi^j_u \, dB^j_u
\ee
represents the portfolio's {\it funding costs}. Such a simple additive decomposition of the wealth process
will no longer hold when more constraints will be imposed on trading.
\erem

\brem\label{rem:cash}
Sometimes (see, e.g., \cite{PV10}), the process $\gamma$, which is given by
\bde
\gamma_t = V_0(\phi,\pA) + F_t(\phi,\pA )+\sum_{i=1}^d  \int _{(0,t]} \xi^i_u \,d\pA^i_u+A_t
\ede
for $t\in [0,T]$, is referred to as the \textsl{cash} process financing the portfolio $\phi$.
In this context, it is important to stress that the equality
\bde
V_t(\phi,\pA) = \sum_{i=1}^d  \int _{(0,t]} \xi^i_u \, dS^i_u+\gamma_t,
\ede
holds but, in general, we have that
\bde
V_t(\phi,\pA) \ne \sum_{i=1}^d  \xi^i_t S^i_t + \gamma_t.
\ede
\erem

\ssc{Elementary Market Model} \lab{secelem}

By the {\it elementary market model} we mean a preliminary framework in which trading in funding accounts $B^i$ and risky assets $S^i$ is unconstrained. This is indeed a fairly simplistic set-up, so its analysis should merely be seen as a first step towards more realistic models with trading constraints. We will argue that explicit formulae for the wealth dynamics under various kinds of trading constraints can be obtained from the basic result, Proposition \ref{prop1.1}, by refining the computations involving the wealth process and funding costs.

Let $V^{\textrm{cld}}(\phi , \pA )$ be the {\it netted wealth} of a trading strategy $( \phi, \pA)$, as given by the following equality
\be \lab{portf2c}
 V^{\textrm{cld}}_t (\phi , \pA ) :=  V_t (\phi , \pA ) - B^0_t \int_{(0,t]} (B^0_u)^{-1} \, d\pA_u.
\ee
It is worth noting that the netted wealth $V^{\textrm{cld}}(\phi , \pA )$ is a useful theoretical construct, rather than a practical concept. On the one hand, the cash flows stream $\pA$ is included in the wealth $V(\phi ,A)$;
on the other hand, the same cash flows stream is formally reinvested in $B^0$ and subtracted from~$V(\phi ,A)$.
We will argue that the discounted netted wealth given by \eqref{portf2c} (or some extension of this formula) will be a convenient tool to examine arbitrage-free property for trading under funding and collateralization.

\brem Obviously, the wealth process depends on a strategy $\phi $ and contractual cash flows $A$,
so that notation $V(\phi , A)$ makes perfect sense. However, for the sake of brevity, the shorthand
notation $V(\phi )$ is used in the remaining part of Section \ref{secfun}.
\erem

We introduce the following notation
\be \lab{portf2a}
 K^i_t :=  \int_{(0,t]} B^i_u \, d\wh S^{i}_u + \pA^i_{t} =  \int_{(0,t]} B^i_u \, d\wh S^{i,\textrm{cld}}_u
\ee
and
\be \lab{portf2b}
K^{\phi }_t :=  \int_{(0,t]} B^0_u \, d\wt V_u (\phi ) - \pA_{t} = \int_{(0,t]} B^0_u \, d\wt V^{\textrm{cld}}_u (\phi )
\ee
where we set $\wt  V^{\textrm{cld}}(\phi ) :=  (B^0)^{-1}V^{\textrm{cld}}(\phi )$ and
$\wt  V(\phi ) :=(B^0)^{-1}  V (\phi )$. Obviously,
\be \lab{hyhy}
\wt  V^{\textrm{cld}}_t(\phi ) = V_0(\phi ) + \int_{(0,t]} (B^0_u)^{-1} \, dK^{\phi }_u .
\ee

\brem
The process $K^i$ is equal to the wealth, discounted by the funding account $B^i$, of a self-financing strategy
in risky security $S^i$ and the associated funding account $B^i$ in which $B^i_t$ units of
the cumulative-dividend price of the $i$th asset is held at time $t$.
\erem

The following proposition is fairly abstract and is primarily tailored to cover the valuation and hedging of an {\it unsecured} financial derivative. We thus mainly focus here on funding costs associated with trading in risky assets. A study of {\it secured} (that is, {\it collateralized}$\, $) contracts is postponed to the next section.
We will argue later on that this result is a good starting point to analyze a wide spectrum of practically appealing situations (see, in particular, Propositions \ref{collssp}, \ref{collssp1} and  \ref{collssp2}). To achieve our goals, it will be enough to impose specific constraints on trading strategies, which will reflect particular market conditions faced by the hedger (such as different lending, borrowing and funding rates) and/or covenants of a contract under study (such as a collateral or close-out payoffs or benefits stemming from defaults).

\bp \lab{prop1.1}
(i) For any  self-financing strategy $\phi$ we have that for every $t \in [0,T]$
\be  \lab{portf3b}
K^{\phi }_t= \sum_{i=1}^d \int_{(0,t]} \xi^i_u \, dK^i_u + \sum_{i=1}^d \int_{(0,t]} ( \psi^i_u B^i_u + \xi^i_u S^{i}_u ) (\wt B^i_u)^{-1}  \, d\wt B^i_u
\ee
where we set $\wt B^i := (B^0)^{-1} B^i$. \hfill \break
(ii) The equality
\be  \lab{portf3bc}
K^{\phi }_t= \sum_{i=1}^d \int_{(0,t]} \xi^i_u \, dK^i_u ,  \quad t\in [0,T] ,
\ee
holds if and only if
\be \lab{conee}
\sum_{i=1}^d \int_{(0,t]} ( \psi^i_u B^i_u + \xi^i_u S^{i}_u ) (\wt B^i_u)^{-1}  \, d\wt B^i_u  = 0 , \quad t\in [0,T].
\ee
(iii) In particular, if for each $i=1,2,\dots ,d$ we have that: either $B^i_t=B^0_t$ for all $t \in [0,T]$ or
\be \lab{portf3a}
\psi^i_t B^i_t + \xi^i_t  S^{i}_t = 0, \quad t\in [0,T],
\ee
then (\ref{conee}) is valid and thus (\ref{portf3bc}) holds. \hfill \break
(iv) Assume that $B^i=B^0$ for every $i=1,2,\dots ,d$ and denote $\wt S^{i,\textrm{cld}} = (B^0)^{-1}S^{i,\textrm{cld}} $. Then
\be \lab{class2}
d\wt V^{\textrm{cld}}_t (\phi )= \sum_{i=1}^d\xi^i_t \, d\wt S^{i,\textrm{cld}}_t .
\ee
\ep

\proof
Recall that  (see (\ref{portf2}))
\bde
dV_t (\phi ) =  \sum_{i=1}^d  \xi^i_t \, d(S^i_t + \pA^i_t) + \sum_{j=0}^d \psi^j_t \, dB^j_t  + d\pA_t .
\ede
Using (\ref{rtf1}),  for the discounted wealth $\wt V (\phi ) = (B^0)^{-1} V (\phi )$ we obtain
\begin{align*}
d\wt V_t (\phi ) &=  \sum_{i=1}^d  \xi^i_t \, d((B^0_t)^{-1}S^{i}_t )
+ \sum_{i=1}^d  \xi^i_t  (B^0_t)^{-1} \, d\pA^i_t + \sum_{i=1}^d \psi^i_t \, d( (B^0_t)^{-1}B^i_t)
+ (B^0_t)^{-1} \, d\pA_t \\  &=  \sum_{i=1}^d  \xi^i_t \, d\wt S^{i,\textrm{cld}}_t +
\sum_{i=1}^d \psi^i_t \, d\wt B^i_t + (B^0_t)^{-1} \, d\pA_t
\end{align*}
where $\wt B^i =(B^0)^{-1} B^i$ and
\bde
\wt S^{i,\textrm{cld}}_t = S^{i}_t (B^0_t)^{-1} + \int_{(0,t]} (B^0_u)^{-1} \, d\pA^i_u
= \wt S^{i}_t + \int_{(0,t]} (B^0_u)^{-1} \, d\pA^i_u.
\ede
Consequently,
\begin{align*}
dK^{\phi}_t &= B^0_t \, d\wt V_t (\phi ) - d\pA_t  =  \sum_{i=1}^d  B^0_t \xi^i_t \, d\wt S^{i,\textrm{cld}}_t +\sum_{i=1}^d B^0_t \psi^i_t \, d\wt B^i_t
\\ &=  \sum_{i=1}^d  B^0_t \xi^i_t \, d( \wh S^{i}_t \wt B^i_t ) + \sum_{i=1}^d  B^0_t \xi^i_t \, (B^0_t)^{-1} d\pA^i_t +\sum_{i=1}^d B^0_t \psi^i_t \, d\wt B^i_t  \\
&=  \sum_{i=1}^d  B^0_t \xi^i_t \wh S^{i}_t \, d\wt B^i_t +
\sum_{i=1}^d  B^0_t \wt B^i_t \xi^i_t \, d\wh S^{i}_t  + \sum_{i=1}^d \xi^i_t \, d\pA^i_t +\sum_{i=1}^d B^0_t \psi^i_t \, d\wt B^i_t  \\
&=\sum_{i=1}^d  B^0_t \xi^i_t \wh S^{i}_t \, d\wt B^i_t +
\sum_{i=1}^d \xi^i_t \, (B^i_t \, d\wh S^{i}_t + d\pA^i_t ) +  \sum_{i=1}^d B^0_t \psi^i_t \, d\wt B^i_t \\
&= \sum_{i=1}^d \xi^i_t \, dK^i_t + \sum_{i=1}^d B^0_t ( \psi^i_t + \xi^i_t \wh S^{i}_t ) \, d\wt B^i_t .
\end{align*}
This completes the proof of part (i). Parts (ii) and (iii) now follow easily.
By combining formulae (\ref{portf2a}) and (\ref{portf3b}), we obtain part (iv). Note that \eqref{class2} is the classic
condition for a market with a single savings account $B^0$.
\endproof

\brem \lab{rem2.6}
Note that equality $B^i=B^0$ (resp. equality (\ref{portf3a})) corresponds to unsecured (resp. secured) funding of the $i$th asset (unsecured funding means that a risky security is not posted as collateral).
In the financial interpretation, condition (\ref{portf3a}) means that at any date $t$ the value of the long or short position in the $i$th risky security should be exactly offset by the value of the $i$th secured funding account.
Although this condition is aimed to cover the case of the fully secured funding of the $i$th risky asset using the corresponding repo rate, it is fair to acknowledge that it is rather restrictive and thus not always not practical.
It is suitable for repo contracts with the daily resettlement, but it does not cover the case of long term
repo contracts.

Note also that if condition (\ref{portf3a}) holds for all $i=1,2,\dots, d$ then the wealth of a portfolio $\phi$ satisfies $V_t (\phi ) = \psi^0_t B^0_t$ for every $t \in [0,T]$. This is consistent with the interpretation that
all gains/losses are immediately invested in the savings account $B^0$. To make this set-up more realistic, we
need, in particular, to introduce different borrowing and lending rates and add more constraints on trading.
\erem

\brem
More generally, the $i$th risky security can be funded in part
using $B^i$ and using $B^0$ for another part, so that condition (\ref{portf3a}) may fail to hold.
However, this case can also be covered by  the model in which condition (\ref{portf3a}) is met
by artificially splitting the $i$th asset into two `sub-assets' that are subject different funding rules.
Needless to say that the valuation and hedging results for a derivative security
will depend on the way in which risky assets used for hedging are funded.
\erem

\sssc{The Dynamics of the Wealth Process}

To obtain more explicit representations for the wealth dynamics, we first prove an auxiliary lemma.
From equality (\ref{ortf2a}), one can  deduce that the increment $dK^i_t$ represents the change
in the price of the $i$th asset net of funding cost. For the lack of the better terminology, we propose to
call $K^i$ the \textit{netted realized cash flow} of the $i$th asset.

\bl \lab{lmm1}
The following equalities hold, for all $t \in [0,T]$,
\be \lab{ortf2a}
K^i_t = S^i_t -S^i_0 +\pA^i_t - \int _{(0,t]} \wh S^i_u \, dB^i_u
\ee
and
\be \lab{ortf2b}
K^{\phi}_t = V_t (\phi) - V_0 (\phi) - \pA_t - \int _{(0,t]}  V_u (\phi) \, dB^0_u .
\ee
\el

\proof
The It\^o formula, (\ref{pri2}) and (\ref{portf2a}) yield
\begin{align} \lab{44rr}
 \int_{(0,t]} B^i_u \, d\wh S^{i,\textrm{cld}}_u
&= \int_{(0,t]} B^i_u \, d\wh S^{i}_u + \pA^i_t
=  B^i_t \wh S^{i}_t - B^i_0 \wh S^{i}_0-
\int_{(0,t]} \wh S^{i}_u \, dB^i_u + \pA^i_t \\
&= S^i_t -S^i_0 + \pA^i_t - \int _{(0,t]} \wh S^i_u \, dB^i_u.  \nonumber
\end{align}
The proof of the second formula is analogous.
\endproof

In view of Lemma \ref{lmm1}, the following corollary to Proposition \ref{prop1.1} is immediate.

\bcor \lab{correx}
 Formula (\ref{portf3b}) is equivalent to the following expressions
\begin{align} \lab{class1}
&d\wt V^{\textrm{cld}}_t (\phi )=  \sum_{i=1}^d\xi^i_t \wt B^i_t \, d\wh S^{i,\textrm{cld}}_t
+ \sum_{i=1}^d \zeta^i_t (B^i_t)^{-1}  \, d\wt B^i_t,
\\ \lab{clacss1} &dV_t (\phi )= \wt V_t (\phi )\, dB^0_t+ \sum_{i=1}^d\xi^i_t B^i_t \,
d\wh S^{i,\textrm{cld}}_t + \sum_{i=1}^d  \zeta^i_t (\wt B^i_t)^{-1}  \, d\wt B^i_t
 +  d\pA_t ,\\
 \lab{portf3c1} &dV_t (\phi) =  \wt V_t (\phi) \, dB^0_t+ \sum_{i=1}^d \xi^i_t \, dK^i_t
 +\sum_{i=1}^d \zeta^i_t (\wt B^i_t)^{-1}  \, d\wt B^i_t + d\pA_t
\end{align}
where $\zeta^i_t := \psi^i_t B^i_t + \xi^i_t S^{i}_t $. Hence the funding costs of $\phi $ satisfy
\be \lab{ff66}
F_t (\phi ) = \int _{(0,t]} \wt V_u(\phi )\, dB^0_u +\int_{(0,t]} \sum_{i=1}^d \zeta^i_u
 (\wt B^i_u)^{-1}  \, d\wt B^i_t - \sum_{i=1}^d  \int _{(0,t]} \xi^i_u \wh S^i_u \, dB^i_u .
\ee
\ecor

%

Since the funding costs depend, in particular, on funding accounts $B^0, \dots , B^d$, we will sometimes
emphasize this dependence by writing $F(\phi ) = F(\phi; B^0, \dots , B^d)$.

\bex \lab{ex1}
Suppose that the processes $B^j,\, j=0,1,\dots ,d$ are absolutely continuous, so that they can be represented as
$dB^j_t = r^j_t B^j_t \, dt $ for some $\gg$-adapted processes $r^j,\, j=0,1,\dots ,d$. Then \eqref{clacss1} yields
\be  \lab{pof3d1}
dV_t(\phi)= r^0_t V_t (\phi) \, dt + \sum_{i=1}^d \zeta^i_t (r^i_t-r^0_t)\, dt
+ \sum_{i=1}^d \xi^i_t \big(dS^i_t- r^i_t S^i_t \, dt + d\pA^i_t  \big) + d\pA_t .
\ee
The last formula implies that
\be  \lab{pof6d1}
dV_t(\phi)=  \sum_{j=0}^d r^j_t \psi^j_t B^j_t \, dt + \sum_{i=1}^d \xi^i_t \big(dS^i_t + d\pA^i_t  \big) + d\pA_t,
\ee
which can also be seen as an immediate consequence of \eqref{portf2}.
In particular, the dynamics of funding costs of $\phi $ are given by
\be \lab{pof8d1}
dF_t(\phi ) =\sum_{j=0}^d r^j_t \psi^j_t B^j_t \, dt .
\ee
\eex

\sssc{Common Unsecured Account} \lab{sec213}

In the remaining part of this section, we will examine the consequences of our general results
for various cases of practical interest.
We first assume that $B^i= B^0$ for $i=1,2,\dots ,k$ for some $k \leq d$.
This means that all unsecured accounts $B^1, \dots ,B^k$ fold down into a single cash account,
denoted as $B^0$, but secured accounts $B^{k+1}, \dots ,B^d$ corresponding to repo rates may vary from one asset to another. Formally, it is now convenient to postulate that $\psi^i=0$ for $i=1,2,\dots ,k$ so that a portfolio $\phi $ may be represented as $\phi = (\xi^1,\dots ,\xi^d, \psi^0, \psi^{k+1},\dots ,\psi^d)$. Hence formula \eqref{rtf1} reduces to
\bde
V_t (\phi ) = \sum_{i=1}^d \xi^i_t S^i_t + \sum_{i=k+1}^d \psi^i_t B^i_t + \psi^0_t B^0_t
\ede
and the self-financing condition  \eqref{portf2} becomes
\bde 
V_t (\phi ) = V_0(\phi )+ \sum_{i=1}^d \int_{(0,t]} \xi^i_u \, d(S^i_u + \pA^i_u )
+ \int_{(0,t]} \psi^0_u \, dB^0_u + \sum_{i=k+1}^d \int_{(0,t]} \psi^i_u \, dB^i_u
 + \pA_t .
\ede
Consequently, equality \eqref{clacss1} takes the following form
\be \lab{oo99}
dV_t (\phi )= \wt V_t (\phi )\, dB^0_t+
\sum_{i=1}^k \xi^i_t B^0_t \, d\wt S^{i,\textrm{cld}}_t
+ \sum_{i=k+1}^d\xi^i_t B^i_t \, d\wh S^{i,\textrm{cld}}_t
+ \sum_{i=k+1}^d  \zeta^i_t (\wt B^i_t)^{-1}  \, d\wt B^i_t
 + d\pA_t
\ee
where we denote
\bde 
\wt S^{i,\textrm{cld}}_t := \wt S^i_t+\int_{(0,t]} (B^0_u)^{-1} \, d\pA^i_u ,\quad t\in [0,T] ,
\ede
where in turn $\wt S^i :=  (B^0)^{-1} S^i$.

\bex \lab{ex2} If all accounts  $B^j,\, j=0,1,\dots ,d$ are absolutely continuous so that,
in particular, $r^i = r^0$ for $i=1,2,\dots ,k$, then
\be \lab{t6u}
dV_t (\phi)= \Big( r^0_t \psi^0_t B^0_t+ \sum_{i=k+1}^d r^i_t \psi^i_t B^i_t \Big)\, dt + \sum_{i=1}^d \xi^i_t \big(dS^i_t + d\pA^i_t  \big) + d\pA_t .
\ee
If, in addition, $\zeta^i_t=0$ for $i=k+1,\dots ,d$ then $V_t (\phi ) = \sum_{i=1}^k\xi^i_t S^i_t + \psi^0_t B^0_t $
and \eqref{t6u} yields
\bde
dF_t(\phi ) = r^0_t \Big( V_t (\phi) - \sum_{i=1}^k \xi^i_t S^i_t \Big) \, dt - \sum_{i=k+1}^d \xi^i_t r^i_t S^i_t \, dt.
\ede
\eex

\ssc{Different Lending and Borrowing Cash Rates}  \lab{sec2.2.1}

We now modify our model by
postulating that the unsecured borrowing and lending cash rates are different. Recall that we denoted by $B^{0,+}$ and $B^{0,-}$ the account processes corresponding to the lending and borrowing rates, respectively.
It is now natural to represent a portfolio $\phi $ as
$\phi = (\xi^1,\dots ,\xi^d,\psi^{0,+}, \psi^{0,-}, \psi^{1},\dots ,\psi^d, )$ where, by assumption,
$\psi^{0,+}_t \geq 0$ and $\psi^{0,-}_t \leq 0$ for all $t \in [0,T]$.
Moreover, since simultaneous lending and borrowing of cash is either precluded or not efficient (if $r^{0,-} \geq r^{0,+}$), we also postulate that $\psi^{0,+}_t \psi^{0,-}_t =0$
for all $t \in [0,T]$. The wealth process of a portfolio $\phi $ now equals
\be\label{vifi}
V_t (\phi ) =  \sum_{i=1}^d \xi^i_t S^i_t + \sum_{i=1}^d \psi^i_t B^i_t + \psi^{0,+}_t B^{0,+}_t + \psi^{0,-}_t B^{0,-}_t ,
\ee
and the self-financing condition reads
\begin{align}   \lab{sfvifi}
V_t (\phi ) =\, \, & V_0(\phi )+ \sum_{i=1}^d \int_{(0,t]} \xi^i_u \, d(S^i_u + \pA^i_u ) + \sum_{i=1}^d \int_{(0,t]} \psi^i_u \, dB^i_u \\
&+ \int_{(0,t]} \psi^{0,+}_u \, dB^{0,+}_u + \int_{(0,t]} \psi^{0,-}_u \, dB^{0,-}_u + \pA_t . \nonumber
\end{align}
It is worth noting that $\psi^{0,+}_t$ and  $\psi^{0,-}_t$ satisfy
\bde
\psi^{0,+}_t = (B^{0,+}_t)^{-1} \Big( V_t (\phi ) -  \sum_{i=1}^d \xi^i_t S^i_t -\sum_{i=1}^d \psi^i_t B^i_t\Big)^+
\ede
and
\bde
\psi^{0,-}_t = - (B^{0,-}_t)^{-1} \Big( V_t (\phi ) -  \sum_{i=1}^d \xi^i_t S^i_t- \sum_{i=1}^d \psi^i_t B^i_t \Big)^-.
\ede

The following corollary furnishes the wealth dynamics in the present set-up.

\bcor \lab{corcc}
(i) Assume that $B^{0,+}$ and $B^{0,-}$ are account processes corresponding to the lending and borrowing rates.
Let $\phi $ be any self-financing strategy such that $\psi^{0,+}_t \geq 0,\, \psi^{0,-}_t \leq 0$ and $\psi^{0,+}_t \psi^{0,-}_t =0$ for all $t \in [0,T]$. Then the wealth process $V(\phi )$, which is given by \eqref{vifi}, has the following dynamics
\begin{align} \lab{oo44}
dV_t (\phi ) = \, \, & \sum_{i=1}^d\xi^i_t B^i_t \, d\wh S^{i,\textrm{cld}}_t
+ \sum_{i=1}^d  \zeta^i_t (B^i_t)^{-1}  \, dB^i_t
 + d\pA_t \nonumber \\
&+ \Big( V_t (\phi ) -  \sum_{i=1}^d \xi^i_t S^i_t -\sum_{i=1}^d \psi^i_t B^i_t\Big)^+  (B^{0,+}_t)^{-1} \, dB^{0,+}_t
\\&- \Big( V_t (\phi ) -  \sum_{i=1}^d \xi^i_t S^i_t -\sum_{i=1}^d \psi^i_t B^i_t\Big)^- (B^{0,-}_t)^{-1}\, dB^{0,-}_t . \nonumber
\end{align}
(ii) If, in addition, $\psi^i_t=0$ for $i=1,\dots ,k$ and $\zeta^i_t=0$ and $i=1,\dots , d$
for all $t \in [0,T]$ then
\begin{align} \lab{oo4n4}
dV_t (\phi ) =\, \, & \sum_{i=1}^k \xi^i_t \, d(S^i_t + \pA^i_t )
 + \sum_{i=k+1}^d \xi^i_t B^i_t \, d\wh S^{i,\textrm{cld}}_t + d\pA_t
\\ &+ \Big( V_t (\phi )  -  \sum_{i=1}^k \xi^i_t S^i_t \Big)^+  (B^{0,+}_t)^{-1} \, dB^{0,+}_t
- \Big( V_t (\phi )  -  \sum_{i=1}^k \xi^i_t S^i_t \Big)^- (B^{0,-}_t)^{-1}\, dB^{0,-}_t . \nonumber
\end{align}
\ecor

\proof
Formula \eqref{oo44} can be derived from \eqref{sfvifi} using also equality (see \eqref{44rr})
\bde
B^i_t \, d\wh S^{i,\textrm{cld}}_t =dS^i_t   - \wh S^i_t \, dB^i_t+ d\pA^i_t .
\ede
We omit the details.
\endproof

\bex \lab{ex3}
Under the assumptions of part (ii) in Corollary \ref{corcc} if, in addition, all account processes $B^i$ for $i=1,k+2, \dots d+2$ are absolutely continuous then \eqref{oo4n4} becomes
\begin{align}  \lab{oo33}
dV_t(\phi ) =\, \, &\sum_{i=1}^k \xi^i_t \big(dS^i_t + d\pA^i_t  \big) +
\sum_{i=k+1}^d \xi^i_t \big(dS^i_t- r^i_t S^i_t \, dt + d\pA^i_t  \big) + d\pA_t
\\&+ r^{0,+}_t  \Big( V_t (\phi) - \sum_{i=1}^k \xi^i_t S^i_t \Big)^+ \, dt
- r^{0,-}_t  \Big( V_t (\phi) - \sum_{i=1}^k \xi^i_t S^i_t \Big)^- \, dt \nonumber
\end{align}
and thus the funding costs satisfy
\bde
dF_t(\phi ) =r^{0,+}_t  \Big( V_t (\phi) - \sum_{i=1}^k \xi^i_t S^i_t \Big)^+ \, dt
- r^{0,-}_t  \Big( V_t (\phi) - \sum_{i=1}^k \xi^i_t S^i_t \Big)^- \, dt
- \sum_{i=k+1}^d  r^i_t \xi^i_t S^i_t \, dt .
\ede
In particular, by setting $k=0$ we obtain
\be \lab{oov33}
dV_t(\phi ) =\sum_{i=1}^d \xi^i_t \big(dS^i_t- r^i_t S^i_t \, dt + d\pA^i_t  \big) + d\pA_t
+ r^{0,+}_t  \big( V_t (\phi)  \big)^+ \, dt
- r^{0,-}_t  \big( V_t (\phi) \big)^- \, dt .
\ee

\eex

\ssc{Trading Strategies under Various Forms of Netting}  \lab{sec2.2}

So far, long and short positions in funding accounts $B^j,\, j=0,1, \dots , d$ were assumed to bear the same interest. This assumption will be now relaxed, so that we will now deal with the extended framework in which the issue of netting long and short positions in risky assets becomes crucial.
Let us first make some comments about the concept of netting of positions and/or exposures.
In general, the concept netting of long and short positions can be introduced at various levels of inclusiveness, from the absence of netting altogether to the most comprehensive case of netting of all positions. For our purposes,
it will be enough to consider the following cases: \hfill \break
(a) the absence of any netting of long/short positions, \hfill \break
(b) only netting of long/short positions for each particular risky asset and its funding accounts, \hfill \break
(c) in addition, netting of long/short cash positions for all risky assets that are funded from a given funding account, \hfill \break
(d) in addition, netting of exposures (and thus also margin accounts) associated with all contracts between the counterparties.

\sssc{Case (a)}  \lab{sec22.2}

To cover the case of the the total absence of any netting of long and short
positions in any asset $S^i$, one can postulate that for all $i=1,\dots ,d$ and $t \in [0,T]$,
\bde
\xi^{i,-}_t S^i_t+ \psi^{i,+}_t B^{i,+}_t = 0 , \quad
\xi^{i,+}_t S^i_t + \psi^{i,-}_t B^{i,-}_t = 0
\ede
where
\bde
\xi^{i,-}_t S^i_t := - (\xi^i_t S^i_t)^{-} \leq 0, \quad
\xi^{i,+}_t S^i_t := (\xi^i_t S^i_t)^{+} \geq 0
\ede
so that $\psi^{i,+}_t \geq 0$ and $\psi^{i,-}_t \leq 0$ for all $t \in [0,T]$. In particular, even if $\xi^{i,+}_t + \xi^{i,-}_t = 0$ for all $t$, meaning that the net
position in the $i$th asset is null at any time, there will be still an incremental cost
of holding open both positions, due to the spread between the accounts $B^{i,+}$ and $B^{i,-}$.
This case is apparently very restrictive and not practical. Hence it will not be analyzed in what follows.

\sssc{Case (b)} \lab{sec2.2.2}

Let us now examine the case (b). To make this set-up non-trivial, we introduce two different accounts, denoted as $B^{i,+}$ and $B^{i,-}$, which are aimed to reflect the funding costs for the $i$th asset.  We now postulate that
\bde
V_t(\phi ) = \psi^{0,+}_t B^{0,+}_t + \psi^{0,-}_t B^{0,-}_t
+ \sum_{i=1}^d ( \xi^i_t S^i_t  + \psi^{i,+}_t B^{i,+}_t + \psi^{i,-}_t B^{i,-}_t )
= \psi^{0,+}_t B^{0,+}_t + \psi^{0,-}_t B^{0,-}_t
\ede
where $\psi^{i,+}_t \geq 0$ and $\psi^{i,-}_t \leq 0$ for $t \in [0,T]$ and, for $i=1,2, \dots ,d$ and $t \in [0,T]$,
\be \lab{xntt6}
\xi^i_t S^i_t + \psi^{i,+}_t B^{i,+}_t +  \psi^{i,-}_t B^{i,-}_t = 0 .
\ee
The netting mechanism can be here interpreted as follows: for the purpose of hedging, it is pointless to hold
at the same time long and short positions in any asset $i$; it is enough to take the net position in the $i$th asset. For example, if a bank already holds the short position in some asset and the need to take the long position
of the same size arises, we postulate that the short position is first closed. One could notice, however,
that this way of trading is not always an optimal from the point of view of minimization of total funding costs.
Note also that condition  \eqref{xntt6} prevents netting of short or long cash positions within assets for which
long and short funding accounts coincide. See also Remark \ref{rem2.6} for general comments regarding condition (\ref{portf3a}), which also apply to condition \eqref{xntt6}.

Since simultaneous lending and borrowing of cash from the funding account $i$
is not allowed (or not efficient if $r^{i,-} \geq r^{i,+}$),
we also postulate that $\psi^{i,+}_t \psi^{i,-}_t =0$ for all $t \in [0,T]$. This implies that
\be \lab{viigy1}
\psi^{0,+}_t = (B^{0,+}_t)^{-1} ( V_t (\phi ))^+, \quad
\psi^{0,-}_t = - (B^{0,-}_t)^{-1} ( V_t (\phi ))^-
\ee
and, for every $i=1,2,\dots , d$,
\be  \lab{viigy2}
\psi^{i,+}_t = (B^{i,+}_t)^{-1} (\xi^i_t S^i_t)^- , \quad \psi^{i,-}_t = - (B^{i,-}_t)^{-1} (\xi^i_t S^i_t)^+.
\ee
The self-financing condition reads
\bde\lab{portvf2}
V_t (\phi ) = V_0(\phi )+ \sum_{i=1}^d \int_{(0,t]} \xi^i_u \, d(S^i_u + \pA^i_u )
+ \sum_{i=0}^d \int_{(0,t]} \psi^{i,+}_u \, dB^{i,+}_u + \sum_{i=0}^d \int_{(0,t]} \psi^{i,-}_u \, dB^{i,-}_u
+ \pA_t .
\ede
Hence the following result is straightforward.

\bcor \lab{cortcc}
Assume that $B^{i,+}$ and $B^{i,-}$ are account processes corresponding to the lending and borrowing rates.
We postulate that $\psi^{i,+}_t \geq 0,\, \psi^{i,-}_t \leq 0$ and $\psi^{i,+}_t \psi^{i,-}_t =0$ for all
$i=0,1, \dots ,d$ and $t \in [0,T]$, and equality (\ref{xntt6}) holds for all $i=1,2,\dots ,d$.
Then the wealth process equals, for all $t \in [0,T]$,
\bde
V_t(\phi ) =  \psi^{0,+}_t B^{0,+}_t + \psi^{0,-}_t B^{0,-}_t
\ede
and the wealth dynamics are
\begin{align}  \lab{oo4v4}
dV_t(\phi ) = \, \, &\sum_{i=1}^d \xi^i_t \, (dS^i_t + d\pA^i_t)
+ \sum_{i=1}^d ( \xi^i_t S^i_t )^-  (B^{i,+}_t)^{-1} \, dB^{i,+}_t
- \sum_{i=1}^d (\xi^i_t S^i_t )^+ (B^{i,-}_t)^{-1}\, dB^{i,-}_t \\
&+( V_t (\phi ) )^+  (B^{0,+}_t)^{-1} \, dB^{0,+}_t
- ( V_t (\phi ))^- (B^{0,-}_t)^{-1}\, dB^{0,-}_t + d\pA_t . \nonumber
\end{align}
\ecor

\brem
When the equality $B^{i,+}=B^{i,-}= B^i$ holds for all $i=1,2, \dots ,d$ then formula  \eqref{oo4v4} can be seen as
a special case of formula  \eqref{oo44} with $\zeta^i_t=0$ for all $i$ (see also dynamics \eqref{oov33}).
\erem

\bex \lab{ex3a} Under the assumptions of Corollary \ref{cortcc} if, in addition, all account processes $B^{i,+}$
and $B^{i,-}$ for $i=0,1, \dots ,d$ are absolutely continuous then \eqref{oo4v4} becomes
 (note that \eqref{oox33} extends \eqref{oov33})
 \begin{align}  \lab{oox33}
dV_t(\phi ) =\, \, &\sum_{i=1}^k \xi^i_t \big(dS^i_t + d\pA^i_t  \big)
+ \sum_{i=1}^d r^{i,+}_t( \xi^i_t S^i_t )^-  \, dt -  \sum_{i=1}^d r^{i,-}_t (\xi^i_t S^i_t )^+ dt
\\ &+   r^{0,+}_t ( V_t (\phi) )^+ \, dt - r^{0,-}_t ( V_t (\phi))^- \, dt + d\pA_t \nonumber
\end{align}
and thus the funding costs satisfy
\bde
dF_t(\phi ) = r^{0,+}_t ( V_t (\phi) )^+ \, dt - r^{0,-}_t ( V_t (\phi))^- \, dt
+ \sum_{i=1}^d r^{i,+}_t( \xi^i_t S^i_t )^-  \, dt
- \sum_{i=1}^d r^{i,-}_t (\xi^i_t S^i_t )^+ dt .
\ede
\eex

\sssc{Case (c)}  \lab{sec2.2.3}

We will examine here a special case of convention (c), which seems to be of some interest in practice.
We now assume that $B^{i,+} = B^{0,+}$ for all $i =1,2, \dots ,d$ and we postulate
that all short cash positions in risky assets $S^1, \dots , S^d$ are aggregated. This means
that all positive cash amounts available, inclusive of proceeds from short-selling of risky assets,
 are included in the wealth and invested in accounts $B^{0,+}$ or $B^{0,-}$.
By contrast, long cash positions in risky assets $S^i$ are assumed to be funded
from respective funding accounts $B^{i,-}$. We thus deal here with the case of the partial netting of positions
across risky assets.

To formally describe the present set-up, we postulate that
\bde 
V_t (\phi ) = \psi^{0,+}_t B^{0,+}_t + \psi^{0,-}_t B^{0,-}_t
+  \sum_{i=1}^d ( \xi^i_t S^i_t  + \psi^{i,-}_t B^{i,-}_t  )
\ede
where, for every $i=1,2, \dots ,d$ and $t \in [0,T]$, the process  $\psi^{i,-}_t$ satisfies
\be  \lab{biigy2}
 \psi^{i,-}_t =  -(B^{i,-}_t)^{-1} (\xi^i_t S^i_t)^+ \leq 0 .
\ee
so that also
\bde 
V_t (\phi ) = \psi^{0,+}_t B^{0,+}_t + \psi^{0,-}_t B^{0,-}_t - \sum_{i=1}^d ( \xi^i_t S^i_t )^-.
\ede
Since, as usual, it is postulated that $\psi^{0,+}_t \geq 0$ and $\psi^{0,-}_t \leq 0$, we obtain
the following equalities
\bde \lab{biigy1}
\psi^{0,+}_t = (B^{0,+}_t)^{-1} \Big( V_t (\phi ) + \sum_{i=1}^d ( \xi^i_t S^i_t )^- \Big)^+, \quad
\psi^{0,-}_t = - (B^{0,-}_t)^{-1} \Big( V_t (\phi ) + \sum_{i=1}^d ( \xi^i_t S^i_t )^- \Big)^-.
\ede
Finally, the self-financing condition is given by the following expression
\begin{align*} 
V_t (\phi )  = \, \, & V_0(\phi )+ \sum_{i=1}^d \int_{(0,t]} \xi^i_u \, d(S^i_u + \pA^i_u )
+  \sum_{i=0}^d \int_{(0,t]} \psi^{i,-}_u \, dB^{i,-}_u \\
&+ \int_{(0,t]} \psi^{0,+}_u \, dB^{0,+}_u + \int_{(0,t]} \psi^{0,-}_u \, dB^{0,-}_u + \pA_t .
\end{align*}
The following result yields the wealth dynamics in the present set-up.

\bcor \lab{cortccd}
Under the present assumptions, the wealth dynamics are
\begin{align}  \lab{cb4v4}
dV_t (\phi )  = \, \, & \sum_{i=1}^d \xi^i_t \, (dS^i_t + d\pA^i_t)
- \sum_{i=1}^d ( \xi^i_t S^i_t )^+  (B^{i,-}_t)^{-1} \, dB^{i,-}_t + d\pA_t \\
&+  \Big( V_t (\phi ) + \sum_{i=1}^d ( \xi^i_t S^i_t )^- \Big)^+ (B^{0,+}_t)^{-1} \, dB^{0,+}_t
 -  \Big( V_t (\phi ) + \sum_{i=1}^d ( \xi^i_t S^i_t )^- \Big)^- (B^{0,-}_t)^{-1}\, dB^{0,-}_t . \nonumber
\end{align}
\ecor

 Note that even if we assume, in addition, that $B^{i,-} = B^{0,-}$ for all $i=1,2, \dots d$, expression \eqref{cb4v4} does not reduce to \eqref{oo44}, since condition \eqref{biigy2} precludes netting of long cash positions across risky assets.

\bex \lab{ex3b} Under the assumptions of Corollary \ref{cortccd} if, in addition, all account processes $B^{i,+}$ and $B^{0,-}$
 are absolutely continuous then \eqref{cb4v4} becomes
\begin{align}  \lab{cbx33}
dV_t(\phi )  = \, \, & \sum_{i=1}^k \xi^i_t \big(dS^i_t + d\pA^i_t  \big)
- \sum_{i=1}^d r^{i,-}_t( \xi^i_t S^i_t )^+  \, dt + d\pA_t
 \\ &+   r^{0,+}_t \Big( V_t (\phi ) + \sum_{i=1}^d ( \xi^i_t S^i_t )^- \Big)^+ \, dt
- r^{0,-}_t \Big( V_t (\phi ) + \sum_{i=1}^d ( \xi^i_t S^i_t )^- \Big)^- \, dt \nonumber
\end{align}
and thus the funding costs satisfy
\bde
dF_t(\phi ) =  r^{0,+}_t  \Big( V_t (\phi ) + \sum_{i=1}^d ( \xi^i_t S^i_t )^- \Big)^+
 \, dt - r^{0,-}_t  \Big( V_t (\phi ) + \sum_{i=1}^d ( \xi^i_t S^i_t )^- \Big)^- \, dt
 - \sum_{i=1}^d r^{i,-}_t(\xi^i_t S^i_t )^+  \, dt .
\ede
\eex

\subsection{Trading Strategies with Collateralization} \lab{seccoll}

In this section, we will address the situation when the hedger enters a contract with contractual cash flows $A$ and either receives or posts the cash collateral, which, we assume, is specified by some stochastic process $\pC $. Let
\be \lab{collss}
\pC_t = \pC_t \I_{\{ \pC_t \geq 0\}} -  \pC_t \I_{\{ \pC_t < 0\}} = \pC^+_t - \pC^-_t
\ee
be the usual decomposition of $\pC_t$ into the positive and negative components. By convention, $\pC^+_t$ stands for the cash value of collateral received, whereas $\pC^-_t$ represents the cash value of collateral posted. The mechanism of posting or receiving collateral is referred to as  {\it margining}.

In Section \ref{seccoll}, we work under the following standing assumptions: \hfill \break
(a) the lending and borrowing cash rates $B^{0,+}$ and $B^{0,-}$ may be identical or they may differ,  \hfill \break
(b) the long and short funding rates for each risky asset are identical: $B^{i,+}=B^{i,-}=B^i$.

Assumption (b) implies that the issue of netting of long and short cash positions in a given risky asset
is only relevant for assets funded from the cash account (i.e., with $\psi^i_t=0$) and this kind of netting is postulated throughout. Of course, the computations related to the funding of collateral can be combined with any convention regarding netting of cash positions.

\sssc{Generic Margin Account}

Let $B^{\pC,+}, B^{\pCC,+},B^{\pC,-}$ and $B^{\pCC,-}$ be strictly positive, continuous processes of finite variation.
It is easy to see that it suffices to account in the dynamics of the wealth process for additional gains
or losses associated with the variations in the margin account through a minor extension of Definition
\ref{ts1}. To this end, we introduce a {\it collateralized trading strategy} $(\phi , \pA , \pC )$ where we set
\be \lab{vty}
\phi = \big( \xi^1,\dots ,\xi^d, \psi^0 ,\dots ,\psi^d, \psi^{\pC,+} , \psi^{\pC,-}, \psi^{\pCC,+}, \psi^{\pCC,-} \big).
\ee
A portfolio $\phi $ is composed of assets $S^i,\, i=1,2,\ldots,d$, the unsecured account $B^0$, the funding accounts
$B^j,\, j=1,2,\ldots,d$, and the collateral accounts $B^{\pC,+}, B^{\pCC,+},B^{\pC,-}$ and $B^{\pCC,-}$.
The goal of the next definition is merely to introduce additional notation used when analyzing collateralization and rehypothecation. More detailed specification  of processes $\psi^{\pC,+} , \psi^{\pC,-}, \psi^{\pCC,+}$ and  $\psi^{\pCC,-}$ and their financial interpretation will be discussed in the foregoing subsections.
For simplicity,  in Definition \ref{ts2} we assume that $B^{0,+}=B^{0,-}=B^0$. This temporary assumption will be later relaxed.

\bd \lab{ts2}
A collateralized trading strategy $(\phi , \pA , \pC )$ with $\phi $ given by (\ref{vty}) is {\it self-financing} whenever its wealth process $V(\phi )$, which is given by the equality
\be\lab{portf1}
V_t (\phi ) {=}  \sum_{i=1}^d \xi^i_tS^i_t + \sum_{j=0}^d \psi^j_tB^j_t + \psi^{\pC,+}_t B^{\pC,+}_t + \psi^{\pC,-}_t B^{\pC,-}_t + \psi^{\pCC,+}_t B^{\pCC,+}_t +\psi^{\pCC,-}_t B^{\pCC,-}_t,
\ee
satisfies, for every $t \in [0,T]$
\begin{align} \lab{porfx}
V_t (\phi )  = \, \, & V_0(\phi )+ \sum_{i=1}^d \int_{(0,t]} \xi^i_u \, d(S^i_u + \pA^i_u )
+  \sum_{j=0}^d \int_{(0,t]} \psi^j_u \, dB^j_u + \pA_t \\
&+ \int_{(0,t]} \psi^{\pC,+}_u\, dB^{\pC,+}_u + \int_{(0,t]} \psi^{\pC,-}_u \, dB^{\pC,-}_u  + \int_{(0,t]}  \psi^{\pCC,+}_u\, dB^{\pCC,+}_u + \int_{(0,t]} \psi^{\pCC,-}_u \, dB^{\pCC,-}_u. \nonumber
\end{align}
\ed

Definition \ref{ts2} is fairly general, so that it can be used to examine various alternative market conventions that either occur or might occur in practice. In Proposition  \ref{collssp}, we will derive more explicit
representation for the wealth dynamics under {\it segregation}, that is, under the assumption of restricted use of cash collateral when the hedger is collateral taker. Subsequently, in Proposition  \ref{collssp1}  we will address the issue of collateral trading with {\it rehypothecation}. Proposition \ref{collssp2}, which deals with the case of partial  rehypothecation,
covers Propositions \ref{collssp} and \ref{collssp1} as special cases and thus the proofs
of Propositions \ref{collssp} and \ref{collssp1} are omitted.

\sssc{Alternative Specifications of Collateral Amount} \lab{xx1}

In market practice, the collateral amount is typically specified in terms of the mark-to-market value of a hedged contract, whose value at time $t$ is henceforth  denoted as $M_t$.  In this case, we can write
\be \lab{ty6}
\pC_t = (1 +\delta^1_t) M_t \I_{\{ M_t > 0\}} - (1+ \delta^2_t) M_t \I_{\{ M_t < 0\}} = (1+\delta^1_t)
 M^+_t - (1+\delta^2_t) M^-_t
\ee
for some {\it haircut processes} $\delta^1$ and $\delta^2$.
In our theoretical framework, the goal is to develop valuation of a contract through its hedging, so that
it seems natural to tie the mark-to-market value to the (yet unknown) value of a contract.
Since the wealth process $V(\phi )$ is aimed to cover future liabilities of the hedger, the stylized `market value' of a contract, as seen by the hedger, coincides with the negative of his wealth.
Consequently, it makes sense to formally identify the mark-to-market value $M$, as seen from the hedger's perspective, with the negative of the wealth process of hedging strategy. If we set $M= -V (\phi )$ then formula (\ref{ty6}) becomes
\be  \lab{ty6v}
\pC_t = \pC_t (\phi ) := (1+\delta^1_t) V^-_t (\phi )  -  (1+ \delta^2_t) V^+_t (\phi ).
\ee
The case of the fully collateralized contract corresponds to equalities $\delta^1_t = \delta^2_t=0$ for all $t$,
which in turn imply that the equality $\pC (\phi ) = - V(\phi )$ holds. The collateral amount in \eqref{ty6v} is seen from the perspective of the hedger, and thus it depends here on hedger's trading strategy $\phi $. Of course, an analogous analysis can be done for the counterparty. However, since market conditions (in particular, funding rates) are typically different for the two parties, it is not likely that their computations of the contract's value (and thus also the collateral amount) will yield the same outcome.

\sssc{Collateral Trading with Segregated Accounts}

The current financial practice typically requires the collateral amounts to be held in segregated {\it margin accounts}, so that the hedger, as collateral taker, cannot use it for purchasing risky assets, but is required to put it in
the account $B^{\pCC,+}$.  In addition,
we assume that if the hedger is a collateral giver then he needs to borrow the required amount from a
predetermined account $B^{\pCC,-}$. Under these assumptions, the right-hand side in \eqref{portf1} should not
explicitly depend on the collateral process. Formally, we postulate that the
following conditions are met, for all $t\in [0,T],$
\be
\psi^{\pC,+}_t B^{\pC,+}_t + \psi^{\pCC,+}_t B^{\pCC,+}_t = 0 , \lab{33a}
\ee
and
\be
 \psi^{\pC,-}_t B^{\pC,-}_t +\psi^{\pCC,-}_t B^{\pCC,-}_t=0. \lab{33b}
\ee
An important practical case, in which the above conditions are satisfied, is described in Proposition \ref{collssp}.
 We assume henceforth that the collateral accounts $B^{\pC,+}, B^{\pCC,+},B^{\pC,-}$ and $B^{\pCC,-}$
 are subject to the following interpretation: \begin{itemize}
 \item If the hedger receives collateral then he pays to the other party interest determined by the level of $\pC^+$ and the account $B^{\pC,+}$ and he invests the collateral amount in the account $B^{\pCC,+}$.
\item If the hedger is required to post collateral then he borrows the collateral amount $\pC^-$ at the interest specified by $B^{\pCC,-}$ and he receives interest payments determined by the level of $\pC^-$ and the account $B^{\pC,-}$.
\end{itemize}

The next proposition is a rather straightforward extension of Proposition \ref{prop1.1}.
Since this result is easy to establish by combining Corollary \ref{correx} with Definition  \ref{ts2},
we omit the proof. Note that equalities \eqref{piii}--\eqref{xpiii} ensure that conditions
\eqref{33a}--\eqref{33b} are indeed satisfied.

\bp \lab{collssp}
Assume that a trading strategy $(\phi , \pA , \pC )$,  with the process $\phi $ given by (\ref{vty}), is self-financing
and the following equalities hold, for every $t \in [0,T]$,
\begin{align}
&\psi^{\pC,+}_t =-  (B^{\pC,+}_t)^{-1}\pC_t^+,\quad \psi^{\pC,-}_t =(B^{\pC,-}_t)^{-1} \pC_t^- , \lab{piii}\\
&\psi^{\pCC,+}_t = (B^{\pCC,+}_t)^{-1}\pC^+_t ,\quad \psi^{\pCC,-}_t = -(B^{\pCC,-}_t)^{-1}\pC^-_t \lab{xpiii} .
\end{align}
Then the wealth process $V(\phi )$ equals, for every $t \in [0,T]$,
\be\lab{portf1v}
V_t (\phi ) =  \sum_{i=1}^d \xi^i_tS^i_t + \sum_{j=0}^d \psi^j_tB^j_t
\ee
and it admits the following decomposition
\be \lab{portfx}
V_t (\phi ) = V_0(\phi ) + G_t(\phi ) + F_t(\phi ) + \pFC_t + \pA_t
\ee
where $G_t(\phi)$ is given by (\ref{gg66}), $F_t(\phi )$ satisfies (\ref{ff66}),
and the funding costs of the margin account, denoted as $\pFC$, equal
\be \lab{funcos}
\pFC_t =  \int_{(0,t]} \pC^+_u \,  \big( (B^{\pCC,+}_u)^{-1} \, dB^{\pCC,+}_u - (B^{\pC,+}_u)^{-1} \, dB^{\pC,+}_u  \big)
 -  \int_{(0,t]} \pC^-_u \,  \big( (B^{\pCC,-}_u)^{-1} \, dB^{\pCC,-}_u - (B^{\pC,-}_u)^{-1} \, dB^{\pC,-}_u \big) . \nonumber
\ee
More explicitly, the dynamics of the wealth process $V(\phi )$ are
\be  \lab{pof3d1x}
dV_t (\phi)  =  \wt V_t (\phi) \, dB^0_t + \sum_{i=1}^d \xi^i_t \, dK^i_t +
 \sum_{i=1}^d  \zeta^i_t (\wt B^i_t)^{-1}  \, d\wt B^i_t + d\pFC_t + d\pA_t .
\ee
In particular, under assumption (\ref{conee}) we obtain
\be  \lab{2i}
dV_t (\phi) =  \wt V_t (\phi) \, dB^0_t+ \sum_{i=1}^d \xi^i_t \, dK^i_t + d\pFC_t + d\pA_t.
\ee
\ep

Let us define the {\it collateral-adjusted cash flows} $\pAC $ by setting
$\pAC = \pA + \pFC $, so that $ V(\phi ) = V_0(\phi ) + G (\phi ) +F (\phi ) + \pAC $.
Although the funding cost of collateral $\pFC $ will typically depend on the choice of a hedging strategy $\phi $, in order to
keep our notation simple, we do not emphasize this dependence explicitly in the notation of $\pFC $ and $\pAC $.
The definitions and results of Section \ref{secfun} remain valid if we replace $\pA $ by $\pAC $,
provided that the wealth process $V(\phi )$ satisfies \eqref{portf1v}, that is, assuming segregation of collateral. A trading strategy $(\phi ,\pA , \pC )$ satisfying Definition \ref{ts2}, as well as satisfying the segregation of collateral requirement,
can thus formally be reduced to the pair $(\phi ,\pAC )$ satisfying Definition \ref{ts1}.

In particular, the cumulative wealth process  $V^{\textrm{cld}}(\phi )$ is defined through
the following modification of formula (\ref{portf2c})
\be \lab{tportf2c}
 V^{\textrm{cld}}_t (\phi ) :=  V_t (\phi ) - B^0_t \int_{(0,t]} (B^0_u)^{-1} \, d\pAC_u.
\ee
Such reduction comes in handy when the collateral process $\pC$ is independent of the choice
of a portfolio $\phi = ( \xi^1,\dots ,\xi^d, \psi^0 ,\dots ,\psi^d)$. In that case, the valuation and hedging
of a derivative security will simultaneously cover the cash flows of the contract and funding costs of collateral.

\bex \lab{ex4}
We place ourselves within the set-up of Example \ref{ex1} and we postulate, in addition, that the processes
$B^{\pC,+}, B^{\pCC,+},B^{\pC,-}$ and $B^{\pCC,-}$ are absolutely continuous, so that
\begin{align*}
&dB^{\pC,+}_t = r^{\pC,+}_t B^{\pC,+}_t \, dt, \quad dB^{\pCC,+}_t = r^{\pCC,+}_t B^{\pCC,+}_t \, dt  ,
\\  &dB^{\pC,-}_t = r^{\pC,-}_t B^{\pC,-}_t \, dt, \quad  dB^{\pCC,-}_t = r^{\pCC,-}_t B^{\pCC,-}_t \, dt ,
\end{align*}
for some interest rate processes $r^{\pC,+}, r^{\pCC,+},r^{\pC,-}$ and $r^{\pCC,-}$. Then
\be \lab{acgmm}
\pFC_t =  \int_{0}^t  (r^{\pCC,+}_u - r^{\pC,+}_u )\pC^+_u  \, du
 - \int_{0}^t   (r^{\pCC,-}_u - r^{\pC,-}_u )\pC^-_u  \, du .
\ee
Suppose that condition (\ref{conee}) is satisfied. Then, from (\ref{2i}), we obtain
\be  \lab{portf3d}
dV_t (\phi) = r^0_t V_t (\phi) \, dt + \sum_{i=1}^d \xi^i_t \,  \big(dS^i_t- r^i_t S^i_t \, dt + d\pA^i_t  \big)  + d\pFC_t + d\pA_t .
\ee
In the special case when  $r^{\pCC,+}=r^{\pCC,-}=r^0$ and $r^{\pC,+}=r^{\pC,-}=r^{\pC}$, formula (\ref{acgmm}) simplifies to
\be \lab{acgvm}
\pFC_t  =   \int_{0}^t   (r^0_u - r^{\pC}_u )\pC_u  \, du
\ee
and thus (\ref{portf3d}) becomes
\be  \lab{pcctf3d}
dV_t (\phi) = r^0_t V_t (\phi) \, dt  + \sum_{i=1}^d \xi^i_t \,  \big(dS^i_t- r^i_t S^i_t \, dt + d\pA^i_t  \big)
+ (r^0_t - r^{\pC}_t )\pC_t \, dt + d\pA_t .
\ee
Recall that the case of the fully collateralized contract corresponds to
 the equality $\pC = \pC (\phi ) = -V (\phi )$. Under this additional assumption, formula \eqref{pcctf3d} reduces to
\be  \lab{pccmm3d}
dV_t (\phi)=r^{\pC}_t V_t(\phi) \, dt + \sum_{i=1}^d \xi^i_t \big(dS^i_t-r^i_t S^i_t \, dt + d\pA^i_t\big) + d \pA_t .
\ee
Consequently, the funding costs inclusive of the gains/losses from the margin account for the fully collateralized
contract, as seen by the hedger, are
\be  \lab{fund}
F_t (\phi) + \pFC_t (\phi ) = \int_0^t r^{\pC}_u V_u (\phi) \, du - \sum_{i=1}^d \xi^i_u r^i_u S^i_u \, du .
\ee
In a more general situation, when $\pC (\phi ) = \alpha V (\phi )$ for some $\gg$-adapted process $\alpha $, we obtain
\be \lab{acgvmh}
\pFC_t (\phi ) =   \int_{0}^t   (r^0_u - r^{\pC}_u )\alpha_u V_u (\phi ) \, du
\ee
and thus the wealth of a partially collateralized contract is governed by the equation
\be  \lab{pccmm3db}
dV_t (\phi) = \big( (1+ \alpha_t ) r^0_t - \alpha_t  r^{\pC}_t \big) V_t (\phi) \, dt + \sum_{i=1}^d \xi^i_t \big(dS^i_t- r^i_t S^i_t \, dt + d\pA^i_t  \big) + d \pA_t .
\ee
Consequently, the total funding costs of a self-financing trading strategy $(\phi , A, \alpha V (\phi ) )$ are
\be  \lab{fundxx}
F_t (\phi) + \pFC_t (\phi ) = \int_0^t \big( (1+ \alpha_u ) r^0_u -
\alpha_u r^{\pC}_u \big) V_u (\phi) \, du - \sum_{i=1}^d \int_0^t \xi^i_u r^i_u S^i_u \, du .
\ee
Note that the set-up considered in this example can also be easily combined with the set-up of Example \ref{ex2}.
\eex

\sssc{Collateral Trading with Full Rehypothecation}

Rehypothecation is a practice where a bank reuses the collateral pledged by its counterparties as collateral for its own borrowing. In our stylized approach to funding effects of rehypothecation,
 it is natural to assume instead that the hedger, when he is a collateral taker, is granted an unrestricted use of the full collateral amount $\pC^+$. Put another way, the collateral received can be seen as an ordinary component of a hedger's trading strategy (of course,  this applies only prior to counterparty's default). As before, the hedger pays interest on the amount $\pC^+$ to the counterparty at the rate determined by the process $B^{\pC ,+}$. Furthermore, we assume that any traded asset can be posted when the hedger is a collateral giver and when he posts collateral then he is entitled to interest payments, as specified by the process $B^{\pC ,-}$. Note that equality \eqref{xtf1v} reflects the fact that the total amount $V^{\pC}_t (\phi ) := V_t(\phi )+ \pC_t$ can now be used for trading in risky assets by the hedger, where $V(\phi )$ stands for the wealth process exclusive of the collateral amount. This feature makes the present situation different from modeling assumptions
considered so far. The result stated below can be deduced from Proposition \ref{collssp2} and thus we omit the proof.

\bp \lab{collssp1}
Assume that a trading strategy $(\phi , \pA , \pC )$  with $\phi $ given by (\ref{vty}) is self-financing
and the following equalities hold:
\be \lab{piix}
\psi^{\pC,+}_t =- \pC_t^+ (B^{\pC,+}_t)^{-1},\quad \psi^{\pC,-}_t = \pC_t^- (B^{\pC,-}_t)^{-1}, \quad \psi^{\pCC,+}_t = \psi^{\pCC,-}_t = 0,
\ee
so that the wealth process $V(\phi )$ satisfies
\be\lab{xtf1v}
V_t (\phi ) =  \sum_{i=1}^d \xi^i_tS^i_t + \sum_{j=0}^d \psi^j_tB^j_t - \pC_t
\ee
or, equivalently,
\be\lab{xuf1v}
V^{\pC}_t (\phi ) =  \sum_{i=1}^d \xi^i_tS^i_t + \sum_{j=0}^d \psi^j_tB^j_t .
\ee
Then $V(\phi )$ satisfies, for every $t \in [0,T]$,
\be \lab{xrtfx}
V_t (\phi ) = V_0(\phi ) + G_t (\phi ) + F_t (\phi ) + \pFC_t + \pA_t
\ee
where $G_t(\phi)$ is given by (\ref{gg66}), $F_t(\phi )$ satisfies (\ref{ff66}) with $\wt V(\phi)$
replaced by $\wt V^{\pC}(\phi) = (B^0)^{-1} V^{\pC}(\phi )$, and the funding costs of collateral are given by
\be \lab{xncos}
\pFC_t  =  \int_{(0,t]} \pC^-_u  (B^{\pC,-}_u)^{-1} \, dB^{\pC,-}_u  -
\int_{(0,t]} \pC^+_u  (B^{\pC,+}_u)^{-1} \, dB^{\pC,+}_u ,
\ee
or, equivalently,
\be  \lab{vd1x}
dV_t (\phi)  =  \wt V^{\pC}_t (\phi) \, dB^0_t + \sum_{i=1}^d \xi^i_t \, dK^i_t +
 \sum_{i=1}^d \zeta^i_t (\wt B^i_t)^{-1}  \, d\wt B^i_t + d\pFC_t + d\pA_t .
\ee
In particular, under assumption (\ref{conee}) the dynamics of $V(\phi )$ are
\be  \lab{v2i}
dV_t (\phi) =  \wt V^{\pC}_t (\phi) \, dB^0_t+ \sum_{i=1}^d \xi^i_t \, dK^i_t + d\pFC_t + d\pA_t.
\ee
\ep

\bex \lab{ex5}
We work here under the assumptions of Proposition \ref{collssp1}. Our goal is to
provide extensions of formulae obtained in Examples \ref{ex3} and \ref{ex4}.
We therefore postulate that the lending and borrowing rates are different (typically, $r^{0,-} \geq r^0 \geq r^{0,+}$).
We also assume that $r^i=r^0$ for $i=1, \dots ,k$ and condition \eqref{portf3a} is satisfied for $i=k+1,\,k+2,\dots ,d$. Then the wealth $V(\phi )$ equals
\bde
V_t (\phi ) = \sum_{i=1}^k \xi^i_t S^i_t + \psi^{0,+}_t B^{0,+}_t + \psi^{0,-}_t B^{0,-}_t - \pC_t
\ede
where, by assumption, $\psi^{0,+}_t \geq 0$ and $\psi^{0,-}_t \leq 0$, and it satisfies
\be \lab{oo55}
dV_t (\phi )= \wt V^{\pC}_t (\phi )\, dB^{0}_t + dF^{V,\pC}_t (\phi)  +
\sum_{i=k}^d\xi^i_t B^i_t \, d\wt S^{i,\textrm{cld}}_t
+ \sum_{i=k+1}^d\xi^i_t B^i_t \, d\wh S^{i,\textrm{cld}}_t
 + d\pFC_t + d\pA_t
\ee
where in turn $\pFC$ is given by \eqref{xncos} and
\be \lab{tt55}
dF^{V,\pC}_t (\phi ) = \Big( V^{\pC}_t (\phi ) - \sum_{i=1}^k \xi^i_tS^i_t\Big)^+ (r^{0,+}_t - r^0_t)\, dt -
\Big( V^{\pC}_t (\phi ) - \sum_{i=1}^k \xi^i_tS^i_t \Big)^- (r^{0,-}_t - r^0_t) \, dt .
\ee
Equivalently, the dynamics of the wealth process are
\begin{align}  \lab{uu77}
dV_t(\phi ) = \, \, & r^{0,+}_t  \Big( V^{\pC}_t (\phi)  - \sum_{i=1}^k \xi^i_t S^i_t \Big)^+ \, dt
- r^{0,-}_t  \Big( V^{\pC}_t (\phi)  - \sum_{i=1}^k \xi^i_t S^i_t \Big)^- \, dt \\
&+ \sum_{i=1}^k \xi^i_t \big(dS^i_t + d\pA^i_t  \big) +
\sum_{i=k+1}^d \xi^i_t \big(dS^i_t- r^i_t S^i_t \, dt + d\pA^i_t  \big) + d\pFC_t +  d\pA_t \nonumber
\end{align}
and thus the funding costs satisfy
\bde
dF_t(\phi ) = r^{0,+}_t \Big( V^{\pC}_t (\phi)  - \sum_{i=1}^k \xi^i_t S^i_t \Big)^+ \, dt
- r^{0,-}_t\Big( V^{\pC}_t (\phi) - \sum_{i=1}^k \xi^i_t S^i_t \Big)^- \, dt
+ \sum_{i=k+1}^d  r^i_t \psi^i_t B^i_t \, dt .
\ede
\eex

\bex
We work under the assumptions of Example \ref{ex5} and we postulate, in addition, that we deal
with a fully collateralized contract, so that $\pC = - V(\phi )$. Then we obtain the following equalities
\bde
 V^{\pC}_t (\phi) = \sum_{i=1}^k \xi^i_t S^i_t + \psi^{0,+}_t B^{0,+}_t + \psi^{0,-}_t B^{0,-}_t = 0
\ede
and
\begin{align*}
dV_t(\phi )  = \, \, & r^{0,+}_t  \Big(  - \sum_{i=1}^k \xi^i_t S^i_t \Big)^+ \, dt
- r^{0,-}_t  \Big(  - \sum_{i=1}^k \xi^i_t S^i_t \Big)^- \, dt \\
& + \sum_{i=1}^k \xi^i_t \big(dS^i_t + d\pA^i_t  \big) +
\sum_{i=k+1}^d \xi^i_t \big(dS^i_t- r^i_t S^i_t \, dt + d\pA^i_t  \big) + d\pFC_t + d\pA_t .
\end{align*}
Consequently, the funding costs are governed by the following equation
\bde
dF_t(\phi ) = r^{0,+}_t \Big(  - \sum_{i=1}^k \xi^i_t S^i_t \Big)^+ \, dt
- r^{0,-}_t\Big(  - \sum_{i=1}^k \xi^i_t S^i_t \Big)^- \, dt
+ \sum_{i=k+1}^d  r^i_t \psi^i_t B^i_t \, dt .
\ede
If we now assume that $r^{0,+}=r^{0,-}= r^0$ then we get
\bde
 \sum_{i=1}^k \xi^i_t S^i_t + \psi^{0}_t B^{0}_t = 0
\ede
and
\bde
dV_t(\phi ) = \sum_{i=1}^k \xi^i_t \big(dS^i_t - r^0_t S^i_t + d\pA^i_t  \big) +
\sum_{i=k+1}^d \xi^i_t \big(dS^i_t- r^i_t S^i_t \, dt + d\pA^i_t  \big) + d\pFC_t + d\pA_t
\ede
so that
\bde
dF_t(\phi ) = - \sum_{i=1}^k \xi^i_t r^0_t S^i_t \, dt
- \sum_{i=k+1}^d \xi^i_t r^i_t S^i_t \, dt
=  r^0_t \psi^0_t B^i_t \, dt + \sum_{i=k+1}^d  r^i_t \psi^i_t B^i_t \, dt.
\ede
Under an additional assumption that $r^{\pC,+}=r^{\pC,-}= r^{\pC}$, we obtain the following expression
\bde
dV_t(\phi ) = r^{\pC}_t V_t (\phi ) \, dt + \sum_{i=1}^k \xi^i_t \big(dS^i_t - r^0_t S^i_t + d\pA^i_t  \big) +
\sum_{i=k+1}^d \xi^i_t \big(dS^i_t- r^i_t S^i_t \, dt + d\pA^i_t  \big) + d\pA_t ,
\ede
which can also be deduced from  \eqref{pccmm3d}. Finally, for $r^{\pC} = r^0$, we get
\bde
dV_t(\phi ) = r^0_t V_t (\phi ) \, dt + \sum_{i=1}^k \xi^i_t \big(dS^i_t - r^0_t S^i_t + d\pA^i_t  \big) +
\sum_{i=k+1}^d \xi^i_t \big(dS^i_t- r^i_t S^i_t \, dt + d\pA^i_t  \big) + d\pA_t ,
\ede
which can be easily identified as a special case of \eqref{portf3c1}.
\eex

\sssc{Collateral Trading with Partial Rehypothecation}

The amount of assets that can be rehypothecated is sometimes capped; we refer to this situation as the {\it partial
rehypothecation}. The next result addresses the general case of the partial rehypothecation
(equivalently, the case of partial segregation) under different lending and borrowing cash rates. It is thus clear that Proposition \ref{collssp2} covers Propositions \ref{collssp} and \ref{collssp1}
as special cases. Note also that formula \eqref{opo44} is a rather straightforward extension of equality \eqref{oo44}.

\bp \lab{collssp2}
Assume that a trading strategy $(\phi , \pA , \pC )$  with $\phi $ given by
\be \lab{vtyx}
\phi = \big( \xi^1,\dots ,\xi^d, \psi^0 ,\dots ,\psi^d, \psi^{0,+}, \psi^{0,-}, \psi^{\pC,+} , \psi^{\pC,-}, \psi^{\pCC,+}, \psi^{\pCC,-} \big)
\ee
is self-financing and the following equalities hold for all $t \in [0,T]$: $\phi^0_t=0$,
\begin{align*}  
&\psi^{\pC,+}_t =-  (B^{\pC,+}_t)^{-1}\pC_t^+,\quad \psi^{\pC,-}_t =(B^{\pC,-}_t)^{-1} \pC_t^- , \\
&\psi^{\pCC,+}_t = (1- \beta_t) (B^{\pCC,+}_t)^{-1}\pC^+_t ,\quad \psi^{\pCC,-}_t = - (1- \gamma_t) (B^{\pCC,-}_t)^{-1}\pC^-_t ,
\end{align*}
for some $\gg$-adapted stochastic processes $\beta $ and $\gamma $, so that the wealth process $V(\phi )$ equals
\be \lab{xtf4v}
V_t (\phi ) =  \sum_{i=1}^d \xi^i_tS^i_t + \sum_{i=1}^d \psi^i_tB^i_t
+ \psi^{0,+}_t B^{0,+}_t + \psi^{0,-}_t B^{0,-}_t - ( \beta_t \pC^+_t - \gamma_t \pC^-_t).
\ee
We assume that $\psi^{0,+}_t \geq 0,\, \psi^{0,-}_t \leq 0$ and $\psi^{0,+}_t \psi^{0,-}_t =0$ for all $t \in [0,T]$
and
\begin{align} \lab{porifx}
V_t (\phi )  = \, \, & \sum_{i=1}^d \xi^i_t \, d(S^i_t + \pA^i_t )
+  \sum_{i=1}^d  \psi^i_t \, dB^i_t + \psi^{0,+}_t \, dB^{0,+}_t + \psi^{0,-}_t \, dB^{0,-}_t + \pA_t \\
&+  \psi^{\pC,+}_t\, dB^{\pC,+}_t + \psi^{\pC,-}_t \, dB^{\pC,-}_t  + \psi^{\pCC,+}_t\, dB^{\pCC,+}_t
+ \psi^{\pCC,-}_t \, dB^{\pCC,-}_t . \nonumber
\end{align}
Then the wealth process $V(\phi )$ satisfies
\begin{align}  \lab{opo44}
 dV_t (\phi)  = \, \, &\sum_{i=1}^d \xi^i_t B^i_t \, d\wh S^{i,\textrm{cld}}_t +\sum_{i=1}^{d} \zeta^i_t (B^i_t)^{-1}  \, dB^i_t  + \psi^{0,+}_t \, dB^{0,+}_t  + d\pA_t
+ \psi^{0,-}_t \, dB^{0,-}_t  \nonumber \\
 & - \pC^+_t  (B^{\pC ,+}_t)^{-1} \, dB^{\pC ,+}_t
+ \pC^-_t (B^{\pC ,-}_t)^{-1} \, dB^{\pC ,-}_t \\
& + (1 - \beta_t) \pC^+_t  (B^{\pCC ,+}_t)^{-1} \, dB^{\pCC ,+}_t
- (1 - \gamma_t ) \pC^-_t (B^{\pCC ,-}_t)^{-1} \, dB^{\pCC ,-}_t  \nonumber
\end{align}
where the processes $\psi^{0,+}$ and $\psi^{0,-}$ are given by the following expressions
\be \lab{vb1}
\zeta^{d+1}_t = \psi^{0,+}_t B^{0,+}_t =\Big( V_t (\phi ) -  \sum_{i=1}^d \xi^i_tS^i_t - \sum_{i=1}^d \psi^i_tB^i_t +  \beta_t \pC^+_t - \gamma_t \pC^-_t\Big)^+
\ee
and
\be \lab{vb2}
\zeta^{d+2}_t = \psi^{0,-}_t B^{0,+}_t = - \Big( V_t (\phi ) -  \sum_{i=1}^d \xi^i_tS^i_t - \sum_{i=1}^d \psi^i_tB^i_t +  \beta_t \pC^+_t - \gamma_t \pC^-_t\Big)^-.
\ee
\ep

\proof
First, we establish equalities \eqref{vb1} and \eqref{vb2} using \eqref{xtf4v} and our assumption: for all $t \in [0,T]$
\bde
\psi^{0,+}_t \geq 0,\quad \psi^{0,-}_t \leq 0 , \quad \psi^{0,+}_t \psi^{0,-}_t =0 .
\ede
 Next, we recall that (see \eqref{44rr})
\bde
B^i_t \, d\wh S^{i,\textrm{cld}}_t =dS^i_t   - \wh S^i_t \, dB^i_t+ d\pA^i_t .
\ede
Formula \eqref{opo44} now follows from \eqref{porifx} by straightforward computations.
\endproof

The following corollary to Proposition \ref{collssp2}  is immediate. Corollary \ref{corbgt} deals with the case when risky assets $S^i$ for $i=1, 2, \dots ,k$ are traded using cash accounts $B^{0,+}$ and $B^{0,-}$,
whereas risky assets $S^i$ for $i=k+1, \, k+2,\dots ,d$ are traded using respective funding accounts $B^i$.
Of course, it is not hard to extend this result to the case of the market model in which we have two funding accounts,
$B^{i,+}$ and $B^{i,-}$, for each risky asset $S^i$ for $i=k+1, \, k+2,\dots ,d$.

\bcor \lab{corbgt}
Under the assumptions of Proposition \ref{collssp2}, we postulate, in addition, that $\psi^i=0$ for $i=1,2,\dots ,k$ and condition (\ref{portf3a}) holds for $i=k+1,\, k+2,\dots ,k$. Then the wealth process $V(\phi )$ satisfies
\begin{align*}
 dV_t (\phi)  = \, \, & \sum_{i=1}^k \xi^i_t \, (dS^i_t + d\pA^i_t) + \sum_{i=k+1}^d \xi^i_t \,
 \big( dS^i_t   - S^i_t (B^i_t)^{-1} \, dB^i_t+ d\pA^i_t \big) + d\pA_t \\
 &+ \Big( V_t (\phi ) -  \sum_{i=1}^k \xi^i_tS^i_t +  \beta_t \pC^+_t - \gamma_t \pC^-_t\Big)^+ (B^{0,+}_t)^{-1} \, dB^{0,+}_t \\
&- \Big( V_t (\phi ) -  \sum_{i=1}^k \xi^i_tS^i_t +  \beta_t \pC^+_t - \gamma_t \pC^-_t\Big)^- (B^{0,-}_t)^{-1} \, dB^{0,-}_t \\
 & - \pC^+_t  (B^{\pC ,+}_t)^{-1} \, dB^{\pC ,+}_t
+ \pC^-_t (B^{\pC ,-}_t)^{-1} \, dB^{\pC ,-}_t \\
& + (1 - \beta_t) \pC^+_t  (B^{\pCC ,+}_t)^{-1} \, dB^{\pCC ,+}_t
- (1 - \gamma_t ) \pC^-_t (B^{\pCC ,-}_t)^{-1} \, dB^{\pCC ,-}_t.
\end{align*}
\ecor

\bex \lab{ex6}
We work here under the assumptions of Corollary \ref{corbgt}.
A more explicit representation for the wealth dynamics is readily available when account processes
are absolutely continuous.

 It is immediate to see that
\begin{align*}
 dV_t (\phi)  = \, \, & \sum_{i=1}^k \xi^i_t \, (dS^i_t + d\pA^i_t) + \sum_{i=k+1}^d \xi^i_t \, (dS^i_t - r^i_tS^i_t \, dt + d\pA^i_t) + d\pA_t \\
 &+ r^{0,+}_t \Big( V_t (\phi ) -  \sum_{i=1}^k \xi^i_tS^i_t  +  \beta_t \pC^+_t - \gamma_t \pC^-_t\Big)^+ \\
  &- r^{0,-}_t \Big( V_t (\phi ) -  \sum_{i=1}^k \xi^i_tS^i_t  +  \beta_t \pC^+_t - \gamma_t \pC^-_t\Big)^- \\
  & - r^{\pC ,+}_t \pC^+_t \, dt + r^{\pC ,-}_t \pC^-_t \, dt
   + (1 - \beta_t)  r^{\pCC ,+}_t \pC^+_t  \, dt - (1 - \gamma_t )r^{\pCC ,-}_t \pC^-_t \, dt .
\end{align*}
We note that when $r^{0,+}_t = r^{0,-}_t = r^0_t$ and $\beta_t = \gamma_t = 0$ for all $t \in [0,T]$ then the formula above
reduces to equation \eqref{portf3d} with $\pFC$ given by \eqref{acgmm}. Moreover, when $\beta_t = \gamma_t = 1$ for
all $t \in [0,T]$ then it coincides with expression \eqref{uu77} with $\pFC$ given by \eqref{xncos}.
It is clear that other important cases are also covered by Proposition \ref{collssp2}.
In particular, we may now set $\pC = \alpha V(\phi )$ for some  $\gg$-adapted stochastic process $\alpha $.
Recall that a partially collateralized contract corresponds to equality
$\pC (\phi ) = \alpha V (\phi )$ for some process $\alpha $ such that $-1 < \alpha < 0$.
One can also introduce the fully collateralized contract with haircuts by postulating that  (see \eqref{ty6v})
\bde
\pC_t (\phi ) = (1+\delta^1_t) V^-_t (\phi )  -  (1+ \delta^2_t) V^+_t (\phi ).
\ede
Finally, it is possible to combine the set-up considered in Proposition \ref{collssp2} with some convention
regarding netting (for instance, the model examined in Section \ref{sec2.2.3}). Needless to say
that a large variety of model assumptions can be studied on a case-by-case basis.
\eex

\subsection{Trading Strategy with Funding Benefit at Default} \lab{secben}

Let us now assume that an investor may default on his contractual obligations before or on the maturity date $T$ of a contract under consideration. In particular, in the case of his default, he will fail to make a full repayment on his unsecured debt, which is formally represented by a negative position in the unsecured cash account $B^{0,-}$. Let $\theta $ be a random time of default and let $R \in [0,1]$ stand for the investor's recovery rate process (assumed to be $\gg$-adapted).
It is now natural to assume that all trading activities will stop at a random horizon date $\theta \wedge T$.
To account for the investor's benefit at default time $\theta$, it suffices to introduce the adjusted
borrowing account $\bar B^{0,-}$ by setting $\bar B^{0,-}_0=1$ and
\be
d\bar B^{0,-}_t = dB^{0,-}_t - B^{0,-}_t (1 - R_t ) \, dH_t
\ee
where we denote $H_t = \I_{\{t \geq \theta \}}$. It is clear that $\bar B^{0,-}_t = B^{0,-}_t$
on the event $\{ \theta > t\}$. Note also that the jump of $\bar B^{0,-}$ at the random time $\theta$ equals
$\Delta \bar B^{0,-}_{\theta } = - (1- R_{\theta } ) B^{0,-}_{\theta }$.
We also replace $\psi^0_t$ by  $\psi^0_{t-}$ in dynamics (\ref{portf2})
in order to make this process $\gg$-predictable. Then the non-negative jump of the wealth process $V(\phi )$, which
is caused by the jump of the process $\bar B^{0,-}$ at the random time $\theta $, is given by the following
expression $\psi^0_{\theta -} \, \Delta \bar B^{0,-}_{\theta } = - (1- R_{\theta } ) \psi^0_{\theta - } B^{0,-}_{\theta }$.
The financial interpretation of this jump is the hedger's benefit at his own default due to the fact that his
debt to the external lender is not repaid in full.

\subsection{Trading Strategy with  Loss at Default} \lab{secloss}

The last step is to describe the loss at the moment of default of either party.
In case of a default of either one of the counterparties prior to or maturity of the contract, the contract is terminated and close-out payments are transferred.
Since the specification of the close-out payment (and thus also the loss of default)
was the topic of numerous papers, we decided not analyze this part of a contract's specification here.
Modeling and arbitrage pricing issues related to the specification of defaults of counterparties, close-out payments, and the
impact of benefits at
defaults on pricing results will be examined in some detail in the second part of this work.

\section{Arbitrage-Free Models and Martingale Measures}\label{Sec2}

The goal of the preceding section was to analyze the wealth dynamics for self-financing strategies
under alternative assumptions about trading, netting and margining rules. In the next step, we will
provide sufficient conditions for the no-arbitrage property of a market model, given the various trading specifications
considered above.

\ssc{Arbitrage Opportunities under Funding Costs} \lab{secfun2}

\noindent{\bf Special case.} We first place ourselves in the elementary set-up of Section \ref{secelem}  with a single cash account $B^0$. Let $\phi $ be an arbitrary self-financing trading strategy. Then formula (\ref{class1}) yields
\be  \lab{prtf3b}
\wt V^{\textrm{cld}}_t (\phi ) = \wt V^{\textrm{cld}}_0 (\phi ) + \sum_{i=1}^d \int_{(0,t]} \xi^i_u \wt B^i_u \, d\wh S^{i,\textrm{cld}}_u  + \sum_{i=1}^d \int_{(0,t]} ( \psi^i_u + \xi^i_u \wh S^{i}_u ) \, d\wt B^i_u .
\ee
Note that dynamics \eqref{class1} of the process $\wt V^{\textrm{cld}}(\phi )$ do not depend on $A$.
In addition, we postulate that $\phi $ is {\it admissible}, so that the discounted cumulative wealth process
$\wt V^{\textrm{cld}}(\phi )$ is non-negative (or at least bounded from below by a constant).
In principle, one may formulate the following general sufficient condition for the arbitrage-free property of the model:
for any self-financing trading strategy $\phi $ there exists a probability measure $\PT^{\phi} $ on $(\Omega , \G_T)$ such that $\PT^{\phi }$ is equivalent to $\P$ and the process $\wt V^{\textrm{cld}} (\phi )$ is a $(\PT^{\phi }, \gg)$-local martingale. Of course, this condition is rather hard to check, in general, and thus it is not practically appealing. We will thus
search for more specific conditions that are relatively easy to verify since they refer to the existence of some
universal martingale measure for a given trading set-up (and perhaps also for a given class of contracts at hand).

To this end, we will need first to re-examine the concepts of an arbitrage-free model and arbitrage price since,
as we will argue in what follows, the classic notions do not apply to the present non-linear framework.
In particular, we show that the study of the arbitrage-free property of a market model cannot be separated from
the study of hedging strategies for a given class of contracts. The reason is that the presence of incoming or
outgoing cash flows associated with a contract have non-additive impact on the dynamics of the wealth process,
and thus also on the final gains or losses from trading.

\brem \lab{ft6}
Obviously, if there exists $B^k \ne B^0$ then
an arbitrage opportunity arises since we may take $\xi^1 = \ldots = \xi^d=0$ and $\psi^j =0$ for every $j$, except for $j=k$. Then we obtain
\bde
\wt V^{\textrm{cld}}_t (\phi ) = \wt V^{\textrm{cld}}_0 (\phi ) + \int_{(0,t]}  \psi^k_u \, d\wt B^k_u
\ede
and thus we see that the existence of a local martingale measure for the process $\wt V^{\textrm{cld}}(\phi )$ is not ensured, in general. It is thus clear that some additional conditions need to be imposed on the class of trading strategies and/or funding rates to ensure that the model is arbitrage-free.
\erem

\noindent{\bf General case.} As clear from Remark \ref{ft6}, the study of self-financing trading strategies under some form
 of mixed funding of risky assets is rather cumbersome, in general, so that we need to do it on a case-by-case basis.
 We may formulate, however, the generic definition of an arbitrage-free market model.
 We now deal with a model in which the borrowing and lending accounts, $B^{0,+}$ and $B^{0,-}$ are different, in general.
 Let us observe that an explicit specification of the discounted netted wealth process $\wt V^{\textrm{cld}}(\phi , A )$ will depend on additional features of the model at hand.
 In the present set-up, the netted wealth is defined by the following extension of formula \eqref{portf2c}
\be \lab{pootf2c}
 V^{\textrm{cld}}_t (\phi , \pA ) :=  V_t (\phi , \pA ) - B^{0,-}_t \int_{(0,t]} (B^{0,-}_u)^{-1} \, d\pA^+_u
 + B^{0,+}_t \int_{(0,t]} (B^{0,+}_u)^{-1} \, d\pA^-_u
\ee
where $\pA = \pA^+ - \pA^-$ is the decomposition of $\pA$ into its increasing and decreasing components.

 In essence, we say that the hedger, who has an initial capital $x$, can produce an arbitrage opportunity using a
 contract $A$ if he can find an admissible strategy $\phi $ such that the netted wealth at time $T$ is non-negative
 and strictly positive with a positive probability. The financial interpretation of the netted wealth at time
 $T$ reads as follows: the hedger who has the initial capital $x$ enters at time 0 a given contract $A$
 and assumes also the virtual opposite position in the same contract. In particular, the additional cash flow at time 0
 equals 0, since the premia cancel out. Next, he select a hedging strategy for the contract and, at the same time, he uses
 external lenders to fund cash flows associated with the opposite position.
 The idea underpinning the next definition is a comparison of the dynamically hedged `long' position in a given contract with the corresponding `short' position in which all outgoing or incoming cash flows are reinvested in unsecured accounts $B^{0,-}$ and $B^{0,+}$.

 \bd \lab{abop0}
A hedger's {\it arbitrage opportunity} associated with a contract $A$ is any trading strategy $\phi $ such
 that the discounted netted wealth process $\wt V^{\textrm{cld}}(\phi , A )$ is bounded
 from below by a constant and the following conditions are satisfied:  $V_T^{\textrm{cld}}(\phi , A ) \geq L_T ( V_0(\phi ) ) $
 and $\P ( V_T^{\textrm{cld}}(\phi , A ) > L_T ( V_0(\phi ) ) ) > 0 $ where we denote
 $L_T (x)  : = x^+ B^{0,+}_T - x^- B^{0,-}_T$.
\ed

\brem
The postulate that the discounted netted wealth process $\wt V^{\textrm{cld}}(\phi , A )$ is bounded
 from below by a constant is merely a technical condition of {\it admissibility}, which is commonly used to ensure that if the process
 $\wt V^{\textrm{cld}}(\phi , A )$ a local martingale under some probability measure then it is
 a supermartingale. Of course, this issue appears even in the simplest case of valuation of options
 in the Black and Scholes model, so it cannot be avoided when dealing with a general continuous-time framework,
 but it is by no means specific to the non-linear pricing examined in the present work.
\erem

\brem
  Note also that the discount factor is left here somewhat unspecified.
 If the constant in Definition \ref{abop0} is set to be zero, so that the netted wealth is non-negative, it suffices to consider the netted wealth without any discounting and thus the problem of the choice of discounting in Definition \ref{abop0} disappears.
 Otherwise, it will depend on the problem and model at hand (see, for instance, Proposition \ref{prparb1}).
\erem

\brem
For simplicity of presentation, we do not introduce explicitly in the right-hand side of \eqref{pootf2c} default times, close-out payment and benefit at default. Hence this form of self-financing condition is suitable for measuring the impact
of funding and collateral, but it should be slightly amended to cover the cash flows at the time of a default.
A suitable extension is straightforward and it will be done in the second part of this research.
\erem

\noindent {\bf Comments.} Since Definition \ref{abop0} departs from the usual way of introducing the concept of an arbitrage opportunity, we will now make some pertinent comments.
 Let $x = V_0(\phi )$ be the initial capital of the hedger. Then the inequality
 $V_T^{\textrm{cld}}(\phi , A ) >  L_T ( V_0(\phi ))$ reads
 \begin{align*}
 V_T(\phi , A ) > x^+ B^{0,+}_T - x^- B^{0,-}_T + B^{0,-}_T \int_{(0,T]} (B^{0,-}_u)^{-1} \, d\pA^+_u
 - B^{0,+}_T \int_{(0,T]} (B^{0,+}_u)^{-1} \, d\pA^-_u
 \end{align*}
 and it is now clear that we are in fact comparing here the outcomes of a fully dynamic hedging with a semi-static funding
 based on unsecured accounts only. It is thus fair to acknowledge that Definition \ref{abop0} is only the first step towards a more general view of arbitrage opportunities that might arise in the context of differing funding costs and credit qualities of a pool
  of potential counterparties.

 Its natural extension would rely on a comparison of two fully dynamically hedged positions
 in $A$ so that we would end up with the following condition: an arbitrage opportunity is a pair $(\phi , \psi)$
 of admissible strategies for the hedger such that $V_0(\phi) = V_0(\psi )$ and
 $V_T(\phi , A ) - V_T(\psi , -A) \geq 0 $ and $\P ( V_T(\phi , A ) - V_T(\psi , -A) >0) > 0$.

 This more general view would mean that an arbitrage opportunity could be created by taking advantage of the presence of (at least) two
 potential counterparties with differing credit qualities. Needless to say that this extension would require to introduce
 at least one more potential counterparty, so that the minimal trading model would
 now include the hedger and his two counterparties. This seems to be a promising avenue for the theoretical research that could be pursued in the future. However, this extended definition would require the possibility of taking opposite positions in an OTC contract with identical features with two different counterparties and this does not seem to be a plausible postulate from the practical perspective.

 The arguments in favor of Definition \ref{abop0} can be summarized as follows: in particular cases of market models
 its implementation is relatively easy, it yields explicit conditions that make financial sense and,
 last but not least, it clarifies and justifies the use of the concept
 of a {\it martingale measure} in the general set-up of a market with funding costs, collateralization and defaults.
 To sum up, although Definition \ref{abop0} is open to criticism, it seems to be an adequate tool to deal with
 the hedging and valuation issues in the current non-linear trading environment.

\ssc{Arbitrage-Free Property} \lab{secvn2}

The concept of an arbitrage-free property can now be introduced either with respect to all contracts $A$ that can
be covered by a particular model or by selecting first a particular class ${\cal A}$ of contracts of our interest. In principle, the arbitrage-free property depends also on the initial capital $x$ of the hedger.
Recall that we denote  $L_T (x)  : =  x^+ B^{0,+}_T - x^- B^{0,-}_T$.

\bd \lab{defarbi}
We say that a market model is {\it arbitrage-free} for the hedger with respect to the class ${\cal A}$
of financial contracts whenever no arbitrage opportunity associated with any contract $A \in {\cal A}$
exists. It other words, for any self-financing strategy $\phi $ and any contract $A \in {\cal A}$ if the
discounted netted wealth process $\wt V^{\textrm{cld}}(\phi , A )$ is bounded from below by a constant then
\be \lab{xx22}
\P \big(  V^{\textrm{cld}}_T (\phi , A ) < L_T ( V_0(\phi ) ) \big) > 0 .
\ee
\ed

\brem
Of course, the situation is not symmetric here, that is, a model in which no arbitrage opportunities for
the hedger exist may still allows for arbitrage opportunities for the counterparty. Even when the market
conditions are exactly symmetrical for both parties, the cash flows of a contract are not symmetrical and
thus the prices for both parties may be different.
\erem

\brem \lab{defr6}
If we assume that $B^{0,+}=B^{0,-}=B^0$ then we obtain the following equivalent condition
$\P \big( \wt V^{\textrm{cld}}_T (\phi , A ) <  V_0 (\phi ) \big) > 0 $.
If we now set $A=0$ then the netted wealth
process $V^{\textrm{cld}}(\phi , 0)$  coincides with the wealth process $V (\phi )$
and thus Definition \ref{defarbi} formally reduces to the classic definition of an arbitrage-free market.
Hence our pricing method agrees with the linear arbitrage pricing theory if no frictions are present
in the market model (or when they do not affect a contract at hand).
\erem

\ssc{Hedger's Arbitrage Prices}

 Let us assume that the market is arbitrage-free in the sense of Definition \ref{defarbi}.
 The next step is to define the range of arbitrage prices of a contract with cash flows $A$.
 Let $x$ be an arbitrary initial capital of the hedger and let $p$ stand for a price of a contract at time 0 for the hedger.
 A positive value of $p$ means that the hedger receives the cash amount $p$ at time 0, whereas a
 negative value of $p$ means that he makes the payment $-p$ to the counterparty at time 0. It is clear from the next
 definition that the price may depend on the hedger's initial capital $x$ and is not unique, in general.

 \bd
 We say that $p$ is a {\it hedger's price} for $A$ whenever for
 any trading strategy $\phi $ with the initial wealth $x+p$, and such that the discounted wealth
 process $\wt V(\phi , A )$ is bounded from below by a constant, we have that either
 \be \lab{vv33a}
 \P \big( V_T (\phi , A ) < L_T (x) \big) > 0
 \ee
 or
  \be \lab{vvy33a}
 \P \big( V_T (\phi , A ) = L_T (x) \big) = 1.
 \ee
 \ed

 The financial interpretation of condition \eqref{vv33a} is that if the hedger has the initial capital $x$ and enters the contract $A$ at the price $p$ then he should not be able to construct an admissible trading strategy $\phi $ with $V_0(\phi )=x+p$
 and such that
 \bde
  \P \big( V_T (\phi , A ) \geq L_T(x) \big) = 1 ,
 \ede
 where the inequality is strict with a positive probability.  In other words, the hedged position in the contract $A$
 should not outperform the cash investment in all states of the world at time $T$.  In practice, the initial capital $x<0$ can be interpreted as the amount of cash borrowed by the trading desk from
 its internal funding desk, which should be repaid with interest $B^{0,-}_T$ at time $T$.
  Therefore, an arbitrage opportunity would mean that the price $p$ is high enough
 to allow the hedger to make profits without any risk.
  Of course, condition \eqref{vvy33a} corresponds to the situation when a contract can be replicated;
  this special case of non-linear pricing technique through solutions to non-linear BSDEs is examined in Section \ref{Sec21}.

\sssc{Martingale Measures for the First Model} \lab{sscfir}

Our next goal is to show that the concept of a martingale measure can still be used as a tool, although it
is now less clear how a martingale measure should be chosen.
In this subsection, we continue the study of the market model introduced in Section \ref{secelem}. We postulate, in addition,
that self-financing trading strategies $\phi $ satisfy condition (\ref{conee}), so that equality (\ref{portf3bc}) holds. As was already mentioned (see Remark \ref{rem2.6}), condition (\ref{conee}) means that repo trades are subject
to the instantaneous resettlement, so that the discounted wealth equals $\wt V_t (\phi , \pA ) := (B^{0}_t)^{-1} V_t(\phi , \pA )$.
Since process $A$ is fixed, we skip it from the notation for $\wt V$ in what follows. Then we have the following result, which closely resembles classic results for market models with a single funding account.

\bp \lab{proarb1}
Assume that there exists a probability measure $\PT$ on $(\Omega , \G_T)$ such that the
processes $\wh S^{i,\textrm{cld}},\, i=1,2, \dots ,d$ are $(\PT , \gg)$-local martingales.
 Then the model of Section \ref{secelem} is arbitrage-free.
\ep

\proof
It suffices to observe that the discounted cumulative wealth $\wt  V^{\textrm{cld}}(\phi )$ satisfies
\begin{align*}
\wt  V^{\textrm{cld}}_t(\phi ) &= \wt  V^{\textrm{cld}}_0(\phi )
+ \int _{(0,t]} (B^0_u)^{-1} \, dK^{\phi}_u =  \wt  V^{\textrm{cld}}_0(\phi )
+\sum_{i=1}^d \int _{(0,t]} (B^0_u)^{-1} \xi^i_u \, dK^i_u
\\ &=\wt V^{\textrm{cld}}_0(\phi )+\sum_{i=1}^d \int _{(0,t]} (B^0_u)^{-1} \xi^i_u B^i_u \, d\wh S^{i,\textrm{cld}}_u .
\end{align*}
Hence the proposition follows from the standard argument, which runs as follows: since
$\wt V^{\textrm{cld}}(\phi )$ is a non-negative (or bounded from below by a constant) local martingale,
it is also a supermartingale under $\PT$, which in turn means that arbitrage opportunities are precluded.
\endproof

 Let us now apply definition in order to describe the set of hedger's prices of $A$.
 Recall that here $B^{0,\pm}=B^0$, and that $V_0(\phi)=x+p$.
 After simple computations, we obtain the following representation for \eqref{vv33a}
 \bde
  \P \bigg( p   +
  \sum_{i=1}^d  \int _{(0,T]} (B^0_u)^{-1} \xi^i_u B^i_u \, d\wh S^{i,\textrm{cld}}_u
  + \int_{(0,T]} (B^{0}_u)^{-1} \, d\pA_u < 0 \bigg) > 0 ,
 \ede
 whereas \eqref{vvy33a} means that equality holds with probability one, that is,
 \bde
  \P \bigg( p   +
  \sum_{i=1}^d  \int _{(0,T]} (B^0_u)^{-1} \xi^i_u B^i_u \, d\wh S^{i,\textrm{cld}}_u
  + \int_{(0,T]} (B^{0}_u)^{-1} \, d\pA_u = 0 \bigg) = 1 .
 \ede
 Note that in this simple set-up the set of arbitrage prices $p$ does not depend on the hedger's initial wealth $x$.

 Assume that $A_t = -X \I_{\{t=T\}}$ and $B^i = B^0$ for every $i=1, \dots , d$. Then we obtain the following
 characterization of the set of arbitrage prices for the hedger: either
 \bde  
 \P \bigg( p + \sum_{i=1}^d  \int _{(0,T]} \xi^i_u \, d\wt S^{i,\textrm{cld}}_u < B^{-1}_T X \bigg) > 0
 \ede
 or
 \bde  
 \P \bigg( p + \sum_{i=1}^d  \int _{(0,T]} \xi^i_u \, d\wt S^{i,\textrm{cld}}_u = B^{-1}_T X \bigg) = 1.
 \ede
 One recognizes here the classic case, namely, the notion of an arbitrage price for the hedger
 as any level of a price $p$ that does not allow for creation of a super-hedging strategy for a claim $X$.

\sssc{Martingale Measures for the Second Model}  \lab{sscsec}

We now consider the set-up introduced in Section \ref{sec2.2.3} with netting of short cash positions.
We assume that $x \geq 0$ and we define the discounted wealth by setting $\wt V^{+}_t (\phi , \pA ) := (B^{0,+}_t)^{-1} V_t(\phi , \pA )$.

\bp \lab{prparb1}
Assume that $r^{0,+}_t \leq r^{0,-}_t$ and $r^{0,+}_t \leq r^{i,-}_t$ for $i=1,2, \dots , d$.
Let us denote
\bde
\wt S^{i,+,{\textrm{cld}}}_t =  (B^{0,+}_t)^{-1}S^i_t + \int_{(0,t]} (B^{0,+}_u)^{-1} \, d\pA^i_u .
\ede
If there exists a probability measure $\PT$ on $(\Omega , \G_T)$ such that the
processes $\wt S^{i,+,\textrm{cld}},\, i=1,2, \dots ,d$ are $(\PT , \gg)$-local martingales
then the model of Section \ref{sec2.2.3} is arbitrage-free.
\ep

\proof
From Corollary \ref{cortccd}, we know that the wealth process satisfies (see formula  \ref{cbx33})
\begin{align*}
dV_t(\phi , \pA )  = \, \, & \sum_{i=1}^k \xi^i_t \big(dS^i_t + d\pA^i_t  \big)
- \sum_{i=1}^d r^{i,-}_t( \xi^i_t S^i_t )^+  \, dt + d \pA_t
 \\ &+   r^{0,+}_t \Big( V_t (\phi , \pA ) + \sum_{i=1}^d ( \xi^i_t S^i_t )^- \Big)^+ \, dt
- r^{0,-}_t \Big( V_t (\phi , \pA ) + \sum_{i=1}^d ( \xi^i_t S^i_t )^- \Big)^- \, dt . \nonumber
\end{align*}
Assume that  $r^{0,+}_t \leq r^{0,-}_t$. Then
\begin{align*}
dV_t(\phi , \pA )  \leq \, \, & \sum_{i=1}^k \xi^i_t \big(dS^i_t + d\pA^i_t  \big)
- \sum_{i=1}^d r^{i,-}_t( \xi^i_t S^i_t )^+  \, dt + d\pA_t
 \\ &+   r^{0,+}_t \Big( V_t (\phi , \pA ) + \sum_{i=1}^d ( \xi^i_t S^i_t )^- \Big)^+ \, dt
- r^{0,+}_t \Big( V_t (\phi , \pA ) + \sum_{i=1}^d ( \xi^i_t S^i_t )^- \Big)^- \, dt \\
= \, \, &r^{0,+}_t  V_t (\phi , \pA )\, dt + \sum_{i=1}^k \xi^i_t \big(dS^i_t + d\pA^i_t  \big) + d\pA_t
- \sum_{i=1}^d r^{i,-}_t( \xi^i_t S^i_t )^+  \, dt +   r^{0,+}_t \sum_{i=1}^d ( \xi^i_t S^i_t )^- \, dt
 \\ \leq \, \, &r^{0,+}_t  V_t (\phi , \pA )\, dt + \sum_{i=1}^k \xi^i_t \big(dS^i_t - r^{0,+}_t S^i_t \, dt + d\pA^i_t \big)
  + d\pA_t
\end{align*}
where the last inequality holds since we assumed that $r^{0,+}_t \leq r^{i,-}_t$.
Consequently, the discounted wealth
$\wt V^{+}_t (\phi , \pA ) := (B^{0,+}_t)^{-1} V_t(\phi , \pA )$
satisfies
\bde
d \wt V^{+}_t (\phi , \pA ) \leq
\sum_{i=1}^k \xi^i_t (B^{0,+}_t)^{-1} \big(dS^i_t - r^{0,+}_t S^i_t \, dt + d\pA^i_t  \big) + (B^{0,+}_t)^{-1} \, d\pA_t
=  \sum_{i=1}^k \xi^i_t \, d\wt S^{i,+,{\textrm{cld}}}_t + (B^{0,+}_t)^{-1} \, d\pA_t .
\ede
Furthermore,
\bde
V^{\textrm{cld}}_t (\phi , \pA ) \leq V_t (\phi , \pA ) - B^{0,+}_t \int_{(0,t]} (B^{0,+}_u)^{-1} \, d\pA_u
\ede
and thus the netted discounted wealth $\wt V^{+,\textrm{cld}}_t (\phi , \pA ) := (B^{0,+}_t)^{-1} V^{\textrm{cld}}_t (\phi , \pA )$ satisfies
\bde
d\wt V^{+,\textrm{cld}}_t (\phi , \pA ) \leq  \sum_{i=1}^k \xi^i_t \, d\wt S^{i,+,{\textrm{cld}}}_t .
\ede
The arbitrage-free property of the model now follows by the usual arguments. Specifically, the initial wealth equals $x \geq 0 $
and thus
\bde
 V_T^{\textrm{cld}}(\phi , A ) \leq  B^{0,+}_T x + B^{0,+}_T \sum_{i=1}^k \int_0^T \xi^i_t \, d\wt S^{i,+,{\textrm{cld}}}_t
\ede
whereas $L_T(x) = B^{0,+}_T x$.  Consequently, either $V_T^{\textrm{cld}}(\phi , A ) = L_T(x)$ or $\P ( V_T^{\textrm{cld}}(\phi , A ) < L_T(x))>0$, so that arbitrage opportunities are precluded.
\endproof

The set of hedger's prices $p$ is now characterized by the following condition: either
\begin{align*}
\P \bigg(& x + p + \sum_{i=1}^k \int _{(0,T]} \xi^i_t \big(dS^i_t + d\pA^i_t  \big)
- \sum_{i=1}^d \int _{(0,T]} r^{i,-}_t( \xi^i_t S^i_t )^+  \, dt + \pA_T - \pA_0
 \\ &+  \int _{(0,T]} r^{0,+}_t \Big( V_t (\phi , \pA ) + \sum_{i=1}^d ( \xi^i_t S^i_t )^- \Big)^+ \, dt
-\int _{(0,T]} r^{0,-}_t \Big( V_t (\phi , \pA ) + \sum_{i=1}^d ( \xi^i_t S^i_t )^- \Big)^- \, dt
 < L_T(x) \bigg) > 0  \nonumber
\end{align*}
or the equality holds in the formula above with probability one.

\subsection{Trading Strategies with Margin Account} \lab{seccol2}

Our next goal is to examine specific features related to the presence of the margin account.
We denote
\bde
\pl^+_t := \int_{(0,t]} \big( (B^{\pCC,+}_u)^{-1} \, dB^{\pCC,+}_u - (B^{\pC,+}_u)^{-1} \, dB^{\pC,+}_u \big)
\ede
and
\bde
\pl^-_t :=  \int_{(0,t]} \big( (B^{\pCC,-}_u)^{-1} \, dB^{\pCC,-}_u - (B^{\pC,-}_u)^{-1} \, dB^{\pC,-}_u \big).
\ede

We place ourselves within the set-up of Section \ref{secelem} and we consider the following two cases: \hfill \break
(A)  the process $\pC$ is independent of the hedger's portfolio $\phi $, \hfill \break
(B) the process $\pC $ depends on the hedger's portfolio $\phi $.

\noindent {\bf Case (A).} If the  collateral process $\pC$ is exogenously predetermined, so it is independent
of the hedger's trading strategy, then we may formally consider the process $\pAC  = \pFC + \pA $ as the full specification of a contract under study. In other words, it is here possible to reduce the valuation and hedging
problem to the case of an unsecured contract with cash flows given by the process $\pAC $.
In view of this argument, it suffices to adjust the definition of the cumulative wealth process
$V^{\textrm{cld}} (\phi , \pAC )$ by setting
\be \lab{portf2gc}
V^{\textrm{cld}}_t (\phi , \pAC ) :=  V_t (\phi , \pAC ) - B^0_t \int_{(0,t]} (B^0_u)^{-1} \, d\pAC_u .
\ee
We do not need to impose here any additional restrictions on processes $\pl^+$ and $\pl^-$, since it suffices to apply
directly Proposition  \ref{proarb1}. By contrast,  Proposition  \ref{proarb2} is used when the collateral process
depends on the hedger's trading strategy $\phi $.

\noindent {\bf Case (B).} The next goal is to extend Proposition  \ref{proarb1} in order to cover the case of an arbitrary collateral process $\pC$, which may possibly depend on a strategy $\phi $ chosen by the hedger, so that $C = C(\phi )$ (see, for instance, formula  (\ref{ty6v}) in  Section \ref{xx1}).
To this end, we postulate that the processes $\pl^+$ and $\pl^-$ are nonincreasing and nondecreasing, respectively.
In the case of absolutely continuous processes $B^{\pC,+}, B^{\pCC,+}, B^{\pC,-}$
 and $B^{\pCC,-}$, these conditions are satisfied provided that $r^{\pC,+} \geq r^{\pCC,+}$ and $r^{\pCC,-} \geq r^{\pC,-}$. Then we obtain the following expression for the dynamics of $\wt  V^{\textrm{cld}}(\phi , \pA )$
\bde  
\wt  V^{\textrm{cld}}_t(\phi , \pA ) =   \wt  V^{\textrm{cld}}_0(\phi , \pA )
+\sum_{i=1}^d \int _{(0,t]} (B^0_u)^{-1} \xi^i_u B^i_u \, d\wh S^{i,\textrm{cld}}_u +
\int_{(0,t]} (B^0_u)^{-1}\pC^+_u \, d\pl^+_u
 -  \int_{(0,t]} (B^0_u)^{-1}\pC^-_u \, d\pl^-_u .
\ede

\bp \lab{proarb2}
Let us consider the model of Section \ref{secelem} under assumption (\ref{conee}).
Assume that there exists a probability measure $\PT$ on $(\Omega , \G_T)$ such that $\PT$ is equivalent to $\P$ and
the processes $\wh S^{i,\textrm{cld}},\, i=1,2, \dots ,d$ are $(\PT , \gg)$-local martingales. If, in addition,
the processes $\pl^+$ and $\pl^-$ are nonincreasing and nondecreasing, respectively,
then the model is arbitrage-free.
\ep

Of course, an analogous result can be formulated for the model of Section \ref{sec2.2.3} by extending Proposition \ref{prparb1}.
Let us summarize the differences between cases (A) and (B). If the collateral process $\pC $ is exogenously given then
the corresponding gains/losses process can be treated as a part of the hedged contract.
If, however, the process $\pC$ depends on $\phi $ then this approach is no longer valid and
$\pC$ should be considered as a part of the wealth process of a hedging strategy. Analogous arguments will be applied to the case of a benefit of default.

\subsection{Trading Strategies with Benefits or Losses at Default} \lab{secben2}

If the random amount of the benefit at default does not depend on a trading strategy then
it can be formally treated as a part of the contract to be valued and hedged. Otherwise,
the situation becomes more delicate and thus it is harder to handle at a general level.
A similar comment applies to the concept of the loss at default.



\section{Replication under Funding Costs and Collateralization} \label{Sec21}


We will now  apply our valuation method to the case of a contract that can be replicated.
 For convenience, we denote by $\pD $ the cumulative dividend paid by an OTC contract after its inception, as seen from the hedger's perspective. It is assumed throughout that $\pD $ is a c\`adl\`ag process of finite variation with $\pD_{0}=0$. The cumulative dividend process accounts for all cash flows associated with a given security, such as `dividends'
 either received or paid after time 0 and before or at the contracts maturity date $T$, including
 the terminal payoff $\Delta \pD_T$ and the close-out payoff at default.

\bex If the the unique cash flow associated with the contract is the terminal payment occurring at time $T$,
denoted as $X$, then the cumulative dividend process for this security takes form
\be \lab{baraa}
\pD_t = X \I_{\{t = T\}}.
\ee
For instance, for the issuer of a European call option, there are no dividend payments and the
 terminal payoff equals $X = - (S_T-K)^+$, so that $\pD_t = - (S_T-K)^+ \I_{\{t = T\}}$.
\eex

In what follows, the prices of OTC contracts will always be defined from the perspective of a hedger.
We consider throughout trading strategies satisfying condition (\ref{conee}).

\bd
We say that a trading strategy $(\phi , \pA , \pC )$ {\it replicates} a contract given by $\pD$ whenever the equality $ \pA_t = \pD_t$ holds for every $t \in [0,T]$ and $V_T(\phi ) = 0$.  If a contract can be
replicated by a trading strategy $(\phi ,\pA , C )$ then the wealth $V(\phi)$ is called the {\it ex-dividend price}
associated with $\phi $ and it is denoted $S (\phi )$. The {\it cum-dividend price} $\Scum $ is defined as
$\Scum_t (\phi ) = S_t (\phi) - \Delta \pD_t$.
\ed

Note that $S_T = V_T(\phi ) = 0$ and $\Scum_T = V_{T-} (\phi )= - \Delta \pD_T $.
In particular, for a call option, we obtain $ \Scum_T = (S_T-K)^+$.

\ssc{General Valuation Results}

Recall that $\pC = \pC (\phi )$, in general. Therefore,
it is not clear whether the uniqueness of the price $S(\phi )$ holds, in the sense that
if $(\phi , \pA, \pC (\phi))$ and $(\wt \phi , \pA , \pC (\wt \phi))$ are two replicating strategies
for a given contract then necessarily $V(\phi) = V(\wt \phi)$.

\sssc{BSDE Approach in the First Model}

We consider the model previously examined in  Section \ref{sscfir} and we work under the assumptions of Proposition \ref{proarb1}. Moreover, we postulate that the collateral process $\pC $ is independent of a hedging strategy $\phi $.
Let us write $\wt \E_t(\cdot ):=\E_{\wt \P}(\, \cdot \, |\, {\cal G}_t)$ where $\PT$ is any martingale measure for the model at hand. It is assumed throughout that random variables
whose conditional expectations are evaluated are integrable. For the sake of brevity, we denote $\pDC = \pFC + \pD $.

\bp
Assume that a derivative security $\pD $ can be replicated by a trading strategy $(\phi, \pA, C).$
Then its ex-dividend price process $S (\phi )$ associated with $\phi $ equals
\be \lab{price}
S_t (\phi ) =- B^0_t \, \wt \E_t\bigg(\int_{(t,T]} (B^0_u)^{-1} \, d\pD^{\pC}_u \bigg),\quad t\in [0,T].
\ee
and the cum-dividend price satisfies
\be \lab{cprice}
\Scum_t (\phi ) =-  B^0_t \, \wt \E_t\bigg(\int_{[t,T]} (B^0_u)^{-1} \, d\pD^{\pC}_u \bigg),\quad t\in [0,T].
\ee
\ep

\proof
Assume that $(\phi , \pA , \pC )$ replicates $\pD $. From  (\ref{2i}), we obtain
\be  \lab{t2i}
d\wt V_t (\phi) =  \sum_{i=1}^d (B^0_t)^{-1} \xi^i_t \, dK^i_t + (B^0_t)^{-1} ( d\pFC_t + d\pD_t ) .
\ee
Since $V_T(\phi )=0$, this yields
\bde
- \wt V_t(\phi ) =  \sum_{i=1}^d \int_{(t,T]}(B^0_u)^{-1} \xi^i_u \, dK^i_u + \int_{(t,T]} (B^0_u)^{-1} \, d\pDC_u
\ede
Since processes $K^i$ are $\PT$-martingales, equality (\ref{price}) follows.
\endproof

To alleviate notation, we will usually write $S$ and $\Scum $ instead of $S (\phi )$ and $\Scum (\phi )$.
The same convention will be also applied to other price processes considered in what follows.
The discounted ex-dividend price process $\wt S$ equals
\be \lab{price1}
\wt S_t=  - \wt \E_t\bigg(\int_{(t,T]} (B^0_u)^{-1} \, d\pD^{\pC}_u \bigg),\quad t\in [0,T].
\ee
 The {\it cumulative-dividend price} is given as
\be
S^{\textrm{cld}}_t:=S_t - B^0_t \int_{(0,t]}  (B^0_u)^{-1} \, d\pD^{\pC}_u,\quad t\in [0,T],
\ee
and the {\it discounted cumulative-dividend price} equals
\be \lab{nhnh}
\wt S^{\textrm{cld}}_t:=\wt S_t - \int_{(0,t]}  (B^0_u)^{-1} \, d\pD^{\pC}_u
= - \wt \E_t\bigg(\int_{(0,T]}  (B^0_u)^{-1} \, d\pD^{\pC}_u \bigg),\quad t\in [0,T].
\ee
From the formula above, it follows immediately that the discounted cumulative-dividend price
$\wt S^{\textrm{cld}}$ is a $\gg$-martingale under $\PT$. Let us introduce the following notation  (see (\ref{portf2a}))
\be \lab{portf2v}
 K_t := S_0 + \int_{(0,t]} B^0_u \, d\wt S_u - \pD^{\pC}_t = S_0 + \int_{(0,t]} B^0_u \, d\wt S^{\textrm{cld}}_u ,
\ee
where the second equality follows from (\ref{nhnh}).
It is clear that $\wt S^{\textrm{cld}}$ is a $\gg$-local martingale under $\wt \P$ if and only
if $K$ is a $\gg$-local martingale under $\PT $.
We refer to the martingale property of $\wt S^{\textrm{cld}}$ as to the {\it multiplicative martingale property},
whereas the martingale property of $K$ is termed the {\it additive martingale property}.
The integration by parts formula yields
\be \lab{ortf2v}
K_t = S_t - \int _{(0,t]} \wt S_u \, dB^0_u - \pD^{\pC}_t  ,\quad t\in [0,T].
\ee

\sssc{BSDE Approach in the Second Model}

We now proceed to the model from Section \ref{sscfir} and we work under the assumptions
of Proposition \ref{prparb1}. In particular, $r^{0,+}_t \leq r^{0,-}_t$ and $r^{0,+}_t \leq r^{i,-}_t$ for $i=1,2, \dots , d$. We postulate the existence of a probability measure $\PT$ on $(\Omega , \G_T)$ such that the
processes $\wt S^{i,+,\textrm{cld}},\, i=1,2, \dots ,d$ are $(\PT , \gg)$-local martingales where
\bde
\wt S^{i,+,{\textrm{cld}}}_t =  (B^{0,+}_t)^{-1}S^i_t + \int_{(0,t]} (B^{0,+}_u)^{-1} \, d\pA^i_u .
\ede
Recall that the wealth process now satisfies
\begin{align*}
dV_t(\phi , \pA )  = \, \, & \sum_{i=1}^k \xi^i_t \big(dS^i_t + d\pA^i_t  \big)
- \sum_{i=1}^d r^{i,-}_t( \xi^i_t S^i_t )^+  \, dt + d \pA_t
 \\ &+   r^{0,+}_t \Big( V_t (\phi ) + \sum_{i=1}^d ( \xi^i_t S^i_t )^- \Big)^+ \, dt
- r^{0,-}_t \Big( V_t (\phi ) + \sum_{i=1}^d ( \xi^i_t S^i_t )^- \Big)^- \, dt . \nonumber
\end{align*}
Let us consider an OTC contract with the dividend process $\pD$. We can now ask the following
question: how to find the least expensive way of contract's replication (or super-hedging)?
More explicitly, we search for a strategy $\phi $ satisfying $\wt V_T(\phi , \pD ) = 0$ with
the minimal initial cost. For brevity, let us represent the dynamics of  $\wt V(\phi , \pD )$ by writing (we denote $S=(S^1,\ldots,S^d)$)
\bde
dV_t(\phi , \pD )  =  \sum_{i=1}^k \xi^i_t \big(dS^i_t - r^{0,+}_t S^i_t \, dt + d\pA^i_t  \big)
+ f(t,\xi_t ,S_t )\, dt + d\pD_t .
\ede
Hence the discounted wealth  $\wt V^{0,+}(\phi , D) := (B^{0,+})^{-1} V^{0,+}(\phi , D)$ satisfies
\bde
d\wt V_t(\phi , \pD )  =\sum_{i=1}^k \xi^i_t \, d \wt S^{i,+,{\textrm{cld}}}_t
-  r^{0,+}_t \wt V_t(\phi , \pD )\, dt
+ (B^{0,+}_t)^{-1} \, f(t,\xi_t ,S_t )\, dt +(B^{0,+}_t)^{-1} \, d\pD_t .
\ede
Informally, our valuation problem can now be intuitively represented as the problem of finding the strategy $\phi=(\xi,\psi) $ satisfying  condition (\ref{conee}), so that equality (\ref{portf3bc}) holds, and such that the  $\xi$ portion of this strategy
minimizes the following expectation
\be
S_0(\phi ) =- \, \wt \E \bigg(\int_{(0,T]} (B^{0,+}_u)^{-1} \, \big(
\wt f(u,\xi_u ,S_u , V_t )\, du + d\pD_u \big) \bigg).
\ee
More precisely, we search for a solution $(Z,\xi )$ to the BSDE
\bde
dZ_t  =\sum_{i=1}^k \xi^i_t \, d \wt S^{i,+,{\textrm{cld}}}_t
+ (B^{0,+}_t)^{-1} \, \wt f(t,\xi_t ,S_t , Z_t )\, dt +(B^{0,+}_t)^{-1} \, d\pD_t
\ede
with the terminal value $Z_T=0$ for which the initial value is minimal.
One can also address the issue of finding the least expensive way of super-hedging by
postulating that $Z_T \geq 0$, rather than $Z_T = 0$.

\ssc{Piterbarg's \cite{PV10} Model}

As a simple illustration of our fairly general non-linear hedging and pricing methodology, we present here a detailed study of the valuation problem previously examined by Piterbarg \cite{PV10}. Following \cite{PV10}, we consider here three funding assets
\bde
B^0_t=e^{\int_0^t r^0_u \, du},\quad B^1_t=e^{\int_0^t r^1_u \, du},\quad B^{\pC}_t=e^{\int_0^t r^{\pC}_u \, du}.
\ede
The spreads $r^1-r^{\pC },\, r^1 -r^0 ,\, r^{\pC}-r^0$ represent the \textit{bases}
between the funding rates, that is, the {\it funding bases}. For simplicity of presentation, we assume that $r^{\pCC,+}=r^{\pCC,-}=r^0$ and $r^{\pC,+}=r^{\pC,-} =r^{\pC}$, but no specific ordering of rates $r^0 , r^1$ and $r^{\pC }$ is postulated
a priori. More general situations where $r^{0,+}\ne r^{0,-},\, r^{\pCC,+} \ne r^{\pCC,-},\, r^{\pC,+} \ne r^{\pC,-}$
can also be handled using our approach.

\sssc{Arbitrage-Free Property and Martingale Measure}

We assume that a stock $S^1$  pays continuously dividends at stochastic rate $\kappa $ and has the (ex-dividend) price dynamics under
the real-world probability $\P$
\bde 
dS^{1}_t = S^{1}_t( \mu_t \, dt+\sigma_t \, dW_t), \quad S^{1}_0>0,
\ede
 where $W$ is a Brownian motion under $\P$. The corresponding dividend process $\pA^1$ is given by
\bde
\pA^1_t = \int_0^t \kappa_u S^{1}_u \, du .
\ede
As usual, we write $\wh S^1_t = (B^1_t)^{-1}S^1_t$ and $\wh S_t^{1,\textrm{cld}}=(B^1_t)^{-1} S_t^{1,\textrm{cld}}$.

\bcor
The price process $S^{1}$ satisfies under $\PT$
\bde 
dS^{1}_t = S^{1}_t\left ((r^1_t- \kappa_t) \, dt+\sigma_t \, d\wt W_t\right ),
\ede
where $\wt W$ is a Brownian motion under $\PT$. Equivalently, the process $\wh S^{1,\textrm{cld}}$ satisfies
\be \lab{xdx}
d\wh S^{1,\textrm{cld}}_t = \wh S^{1,\textrm{cld}}_t \sigma_t \, d\wt W_t .
\ee
The process $K^1$ satisfies
\be \lab{kk1}
dK^1_t = dS^{1}_t - r^1_tS^{1}_t \, dt + \kappa_t S^{1}_t \, dt = S^{1}_t\sigma_t \, d\wt W_t
\ee
and thus it is a (local) martingale under $\wt \P$.
\ecor

\proof By the definition of a martingale measure $\PT$, the discounted cumulative-dividend price  $\wh S^{1,\textrm{cld}}$
is a $\PT$-(local) martingale. Recall that the process $\wh S^{1,\textrm{cld}}$ is given by
\bde  
\wh S^{1,\textrm{cld}}_t = \wh S^1_t+\int_{(0,t]} (B^1_u)^{-1} \, d\pA^1_u ,\quad t\in [0,T].
\ede
Consequently,
\bde
\wh S^{1,\textrm{cld}}_t = \wh S^{1}_t+\int_{(0,t]} \kappa_u (B^1_u)^{-1} S^{1}_u \, du =  \wh S^{1}_t+\int_{(0,t]} \kappa_u \wt S^{1}_u \, du .
\ede
Since
\be\lab{S1x}
d\wh S^{1}_t = \wh S^{1}_t\big( (\mu_t - r^1_t) \, dt+\sigma_t \, dW_t\big),
\ee
we obtain
\bde
d \wh S^{1,\textrm{cld}}_t = d \wh S^{1}_t + \kappa_t \wh S^{1}_t \, dt
= \wh S^{1}_t\big( (\mu_t + \kappa_t - r^1_t) \, dt+\sigma_t \, dW_t\big).
\ede
Hence $\wh S^{1,\textrm{cld}}$ is a $\PT$-martingale provided that the process
\be \lab{ff2}
d\wt W_t =   dW_t + \sigma^{-1}_t (\mu_t + \kappa_t - r^1_t) \, dt
\ee
is a Brownian motion under $\PT$. By combining (\ref{S1x}) with (\ref{ff2}) we obtain expression (\ref{xdx}).
Other asserted formulae now follow easily.
\endproof

\sssc{Valuation of a Collateralized Derivative Security}

\noindent {\bf Exogenous collateral.} Our first goal is to value and hedge a collateralized security with a bounded terminal payoff $X$ at time $T$ and a predetermined collateral process $\pC$.  Note that here $\pD $ is given by formula (\ref{baraa}).

To this end, we consider an admissible trading strategy $\phi = (\xi^1, \psi^0 , \psi^1 , \pD , \pC )$  composed of a dividend-paying stock $S^1$,
the unsecured funding account $B^0$ and the funding account $B^1$. The wealth process $V(\phi )$ is given by the formula
\bde 
V_t (\phi ) :=  \xi^1_tS^1_t + \psi^0_t B^0_t+  \psi^1_tB^1_t
\ede
for every $t \in [0,T]$. We assume that  $ \xi^1_tS^1_t+\psi^1_tB^1_t=0$ for every $t \in [0,T]$ (hence condition (\ref{conee}) is satisfied) so that $V_t (\phi ) = \psi^0_t B^0_t$. Using (\ref{pcctf3d}), we obtain
\be  \lab{pcctf3db}
dV_t (\phi) = r^0_t V_t (\phi) \, dt  + \xi^1_t \, dK^1_t + (r^0_t - r^{\pC}_t )\pC_t \, dt + d \pD_t
\ee
where $\pD_t = X \I_{\{t = T\}}$.

 In view of (\ref{kk1}) and (\ref{pcctf3db}), the discounted wealth satisfies
\be  \lab{pcctf3dm}
d\wt V_t (\phi) =(B^0_t)^{-1} \xi^1_t  S^{1}_t\sigma_t \, d\wt W_t
+ (B^0_t)^{-1}(r^0_t - r^{\pC}_t )\pC_t \, dt + (B^0_t)^{-1} d \pD_t.
\ee
Under the assumption that
\be \label{assume}
V_T(\phi )=0,
\ee
the process  $V_t (\phi )$ coincides with the ex-dividend price $S$ of the
contract with the cumulative dividend $\pD_t = X \I_{\{t = T\}}$ and collateral $\pC$.

\brem
Note that (\ref{pcctf3db}) can be rewritten as follows
\bde \lab{ccpp2}
dV_t (\phi) = \Big( - r^{\pC } \pC_t + r^0_t ( V_t (\phi) + \pC_t) - r^1_t \xi^1_t S^1_t + \kappa \xi^1_t S^1_t \Big) dt + \xi^1_t \, dS^1_t + d\pD_t .
\ede
Upon setting, $\pC (t) = - \pC_t$, the dynamics of $V(\phi )$ become
\bde \lab{ccpp2x}
dV_t (\phi ) = \Big( r^{\pC } \pC (t) + r^0_t ( V_t (\phi) - \pC (t)) - r^1_t \xi^1_t S^1_t + \kappa \xi^1_t S^1_t \Big) dt + \xi^1_t \, dS^1_t + d\pD_t .
\ede
This coincides with the formula derived in Piterbarg \cite{PV10}, although the term $dD_t$ does not appear in \cite{PV10} due to the fact that the cum-dividend wealth process is considered therein.

Consequently, the cash process $\gamma$ (cf. Remark~\ref{rem:cash}) satisfies
\beq\label{eq:cash}
d\gamma_t & =& \Big( - r^{\pC } \pC_t + r^0_t ( V_t (\phi) + \pC_t) - r^1_t \xi^1_t S^1_t + \kappa \xi^1_t S^1_t \Big) dt + d\pD_t \nonumber \\
&=& \Big( r^{\pC } \pC (t) + r^0_t ( V_t (\phi) - \pC (t)) - r^1_t \xi^1_t S^1_t + \kappa \xi^1_t S^1_t \Big) dt + d\pD_t,
\eeq
which was already observed in Piterbarg \cite{PV10} (modulo the absence of the term $dD$ in his equation for $d\gamma $).
\erem

\bp \lab{prox4}
A collateralized contract with the cumulative dividend $\pD_t = X \I_{\{t = T\}}$ and the predetermined collateral process $\pC$ can be replicated by
an admissible trading strategy. Moreover,  the ex-dividend price process satisfies, for  every $t < T$,
\be\lab{K3}
S_t =  - B^0_t \, \wt \E_t \bigg( (B^0_T)^{-1} X  + \int_t^T (B^0_u)^{-1} ( r^0_u - r^{\pC}_u )\pC_u \, du \bigg).
\ee
Equivalently,
\be \lab{K3T}
S_t =  - B^{\pC}_t \, \wt \E_t \bigg( (B^{\pC}_T)^{-1} X  + \int_t^T (B^{\pC}_u)^{-1} ( r^0_u - r^{\pC}_u )(\pC_u+V_u(\phi)) \, du \bigg).
\ee
\ep

\proof
Formula (\ref{K3}) is an immediate consequence of (\ref{price}). The component $\xi^1$ of the replicating strategy is
derived by noting that from (\ref{pcctf3dm}), we obtain
\bde
- (B^0_T)^{-1} X - \int_0^T (B^0_t)^{-1}  (r^0_t - r^{\pC}_t ) \pC_t \, dt - V_0 (\phi ) =
\int_0^T  \xi^1_t (B^0_t)^{-1} S^{1}_t \sigma_t \, d\wt W_t .
\ede
Next, we set $\psi^0_t =(B^0_t)^{-1} V_t (\phi )$ and $\psi^1_t = - (B^1_t)^{-1} \xi^1_tS^1_t$.
To obtain (\ref{K3T}), it suffices to observe that equation \eqref{pcctf3db} can be written as
\be  \lab{pcctf3db-1}
dV_t (\phi) = r^{\pC}_t V_t (\phi) \, dt  + \xi^1_t \, dK^1_t + (r^0_t - r^{\pC}_t )(\pC_t+V_t(\phi)) \, dt + d \pD_t
\ee
and apply the same argument as above.
\endproof

\brem Observe that equivalence of formulae \eqref{K3} and \eqref{K3T} indicates that the choice of discount factor can be rather arbitrary, as long as security's (cumulative) cash flow process is appropriately modified. In case of formula \eqref{K3} the discount factor is chosen as the price process $B^0$ representing a traded asset, whereas in case of formula \eqref{K3T} the discount factor is chosen as the process $B^C$, which is not even a traded asset in the present set-up. Note, in particular, that none of the two choices of the discount factor correspond to the spot martingale measure $\widetilde \P$ which, in the case of dividend rate $\kappa =0$, corresponds to the choice of $B^1$ as the discount factor. In  Section \ref{P}, we provide a more extensive discussion of the above observations in the context of the pricing approach adopted the paper by Pallavicini et al. \cite{PPB12}.
\erem

\newpage

\noindent {\bf Hedger's collateral.} As already mentioned in Section \ref{xx1}, the collateral amount $\pC$ can be specified in terms of the mark-to-market value of a hedged security and thus, at least in theory, it can be given in terms of the wealth process $\phi$ of a hedging strategy. For instance, it may be given as follows (see (\ref{ty6v}))
\be \lab{cc3c}
\pC_t (\phi) = (1+ \delta^1_t) V^-_t (\phi )  - (1+ \delta^2_t)  V^+_t (\phi )
\ee
for some processes $\delta^+$ and $\delta^-$. Consequently, the discounted wealth of a self-financing
strategy satisfies
\be  \lab{pcbbdm}
d\wt V_t (\phi) = \xi^1_t \, dK^1_t + (B^0_T)^{-1}(r^0_t - r^{\pC}_t )
( \delta^+_t V^-_t (\phi )  -  \delta^-_t  V^+_t (\phi ) ) \, dt + (B^0_t)^{-1} d \pD_t .
\ee
We are in a position to formulate the following result.

\bp
The backward stochastic differential equation
\be  \lab{pcxxdm}
dY_t = \xi^1_t \, S^{1}_t\sigma_t \, d\wt W_t + (B^0_T)^{-1}(r^0_t - r^{\pC}_t )
( \delta^+_t Z^-_t  -  \delta^-_t  Z^+_t ) \, dt , \quad Z_T = - X ,
\ee
has the unique solution $(Y, \xi^1)$, where the process $Z$ satisfies
\be \lab{K3cc}
Z_t  = - B^0_t \, \wt \E_t \bigg( (B^0_T)^{-1} X + \int_t^T (B^0_u)^{-1} ( r^0_u - r^{\pC}_u )
( \delta^+_u Z^-_u   -  \delta^-_u  Z^+_u ) \, du \bigg).
\ee
Then the collateralized contract with the cumulative dividend $\pD_t = X \I_{\{t = T\}}$
and the collateral process $\pC$ given by (\ref{cc3c}) can be replicated by
an admissible trading strategy and the ex-dividend price $S_t (\phi )$ equals $Z$ for every $t < T$.
\ep

\proof
One can prove that the price is unique, that is, it does not depend on hedging strategy.
This follows from the general theory of BSDEs with Lipschitz continuous coefficients.
\endproof

\bex Recall that the case of the fully collateralized contract corresponds to the equality
$\pC (\phi) = -V (\phi )$. Under this assumption, we obtain
\bde  
dV_t (\phi) = r^{\pC}_t V_t (\phi) \, dt + \xi^1_t \, dK^1_t + d \pD_t .
\ede
Consequently, assuming \eqref{assume},  we obtain the following BSDE, for $t \in [0,T]$,
\bde  
dV_t (\phi) = r^{\pC}_t V_t (\phi) \, dt + \xi^1_t \, dK^1_t + d \pD_t, \quad V_T(\phi)=0 .
\ede
This also means that $V_t(\phi ) = Z_t$ for all $t < T$, where $Z$ satisfies
\bde  
dZ_t = r^{\pC}_t Z_t \, dt + \xi^1_t \, dK^1_t , \quad Z_T =- X.
\ede
The unique solution to this BSDE equals, for all $t \in [0,T]$,
\be   \lab{K3xx}
Z_t =  - B^{\pC}_t \, \wt \E_t \big( (B^{\pC}_T)^{-1} X \big) = S_t .
\ee
Note that the last equality also follows immediately from \eqref{K3T}. It is interesting to remark that the pricing
formula \eqref{K3xx} combines the expectation under the martingale measure corresponding with discounting
of the stock price using the process $B^1$ with discounting of the cash flow using the process $B^C$, although it
is a special case of a general formula  \eqref{K3cc} where cash flows are discounted using $B^0$.
This illustrates the fact that the choice of a discount factor and a martingale measure, although not completely
arbitrary, but subject to well known rules stemming from the Bayes formula and the It\^o formula, is also to a large extent a matter of a convenient representation of the solution to the valuation problem, rather than the way of defining the price of a contract.
Hence the question about the universal choice of a num\'eraire used for discounting of future cash flows is not well posed,
although for practical purposes such a choice may be beneficial.
\eex

\newpage

\sssc{An Extension}

Let us conclude this work, by making some comments on a more general version of Piterbarg's model.
Suppose that we no longer assume that the equality $\psi^1_tB^1_t + \xi^1_tS^1_t=0$ holds for every $t \in [0,T]$.
From formula (\ref{clacss1}), we see that we need to adjust $\pFC $ to
\be
\wh{F}^C_t := \pFC_t - \int_0^t ( r^1_u - r^0_u) (\psi^1_u B^1_u + \xi^1_u S^1_u )  \, du
\ee
and thus the wealth dynamics become
\bde
dV_t (\phi) = r^0_t V_t (\phi) \, dt + (r^0_t - r^{\pC}_t ) \pC_t \, dt
 + (r^1_t - r^0_t) ( \psi^1_t B^1_t + \xi^1_t S^{1}_t )  \, dt + \xi^1_t \, dK^1_t
\ede
where we assume, for simplicity, that the collateral process $C$ is given.
Consequently, the pricing formula (\ref{K3}) for the claim $X$ takes the following form
\bde 
V_t (\phi ) = B^0_t \, \wt \E_t \bigg( (B^0_T)^{-1}X  - \int_t^T (B^0_u)^{-1} \big( (r^0_u - r^{\pC}_u)\pC_u
+ (r^1_u - r^0_u) ( \psi^1_u B^1_u + \xi^1_u S^{1}_u ) \big) \, du \bigg)
\ede
and the replicating strategy is determined by the equality
\begin{align*}
 (B^0_T)^{-1}X &- \int_0^T (B^0_t)^{-1}  (r^0_t - r^{\pC}_t ) \pC_t \, dt - V_0(\phi )-
  \int_0^T (r^0_t - r^1_t) ( \psi^1_t B^1_t + \xi^1_t S^{1}_t )  \, dt \\ &=
\int_0^T  \xi^1_t B^1_t (B^0_t)^{-1} \, \wt S^{1,\textrm{cld}}_t \sigma_t \, d\wt W_t .
\end{align*}
Let us first assume that $\xi^1_t >0$. Then condition $\psi^1_tB^1_t + \xi^1_tS^1_t > 0$ means that positions in stocks
are partially funded by the unsecured account $B^0$. Therefore, assuming that the inequality $r^0 > r^1$ is satisfied,
 the value of the replicating portfolio
will be now higher than when the hedge is done under the assumption that $\psi^1_tB^1_t + \xi^1_tS^1_t = 0$.

By contrast, the inequality $\psi^1_tB^1_t + \xi^1_tS^1_t < 0$ means
that we are allowed to borrow more cash funded with the account $B^i$ than it is justified by the amount of stock posted as collateral. If the inequality $r^0 > r^1$ holds then the value of the replicating portfolio
will now be lower with respect to the situation when one hedges under our standard assumption that $\psi^1_tB^1_t + \xi^1_tS^1_t = 0$.

In a general case of unrestricted hedging, one faces the problem of solving a suitable optimization problem in order to find the least expensive way of hedging -- this challenging issue is left for the future research.

\ssc{Pallavicini et al. \cite{PPB12} Approach}\label{P}

In Pallavicini et al. \cite{PPB12}, the authors formally introduce the risk-free short-term interest rate
as an `instrumental variable' without assuming that this rate corresponds to a traded asset.
Nevertheless, they start by postulating the existence of a `martingale measure' associated with
discounting of prices of traded assets using this virtual risk-free rate. More importantly,
they also  postulate that the price of any contract can be defined as the conditional expectation
of `discounted cash flows with costs' using this martingale measure (see formula (1) in \cite{PPB12}).

Of course, this valuation recipe cannot be true if applied directly to cash flows of a given contract
without making first some adjustments to cash flows, in order to account for the actual funding costs, margin account,
closeout, etc..  For instance, to deal with the actual funding costs, the authors propose to use formula (17) in \cite{PPB12} as a plausible valuation tool. All formulae on pages 1--26 in in \cite{PPB12} are definitions describing the actual
or  modified cash flows, rather than pricing results derived from fundamentals; we will henceforth focus
on the most intriguing result from \cite{PPB12}, namely, Theorem 4.3. As it is shown in
this result, by changing the probability measure one can avoid using the risk-free
rate and thus the term `instrumental variable' attributed to the risk-free rate seems to be justified.

\newpage

However, as we will argue below, the approach proposed in \cite{PPB12} is somewhat artificial, since it requires a
right guess how to make a suitable cash flows adjustment. More importantly, this rather complicated method is in fact
not needed at all, since it is always enough to focus directly on the right market model with the actual funding costs
and do not postulate any specific shape of the `risk-neutral pricing formula'.

To explain the rationale behind Pallavicini et al. \cite{PPB12} approach, let us consider a market model with a non-dividend paying stock $S^1$ and a savings account $B^0$ such that $dB^0_t = r^0_t B^0_t \, dt$. Although dividends, margin account and closeout can also be covered by the foregoing analysis, for simplicity of presentation, we focus here on the funding costs only.

\bhyp \lab{ahh1}
We assume that our model is arbitrage-free, so that the martingale measure $\PS$ for the process $\wt S^1 = S^1/B^0$ exists.
\ehyp

Let $V (\phi )$ be the wealth of a self-financing trading strategy $\phi = (\xi^1 , \psi^0)$.
The following lemma is well known.

\bl \lab{oop}
The discounted wealth process $\wt V(\phi ) = V(\phi )/B^0$ satisfies the equality $d\wt V_t (\phi ) = \xi^1_t \, d\wt S^1_ t$
and thus it is a $\PS$-local martingale (or a $\PS$-martingale under suitable integrability assumptions).
\el

Let us now define a completely arbitrary process of finite variation, say $B^{\gamma },$ such that
\bde
 dB^{\gamma}_t = \gamma_t B^{\gamma }_t \, dt,\quad t\geq 0,\quad B^\gamma_0>0 .
\ede
It is crucial to stress that it is not postulated that this process represents a traded
asset (or even has anything to do with the market model at hand).
Nevertheless, it still makes sense to make the following assumption.

\bhyp \lab{ahh2}
There exists a probability measure $\P^{\gamma }$ such that the process
$\bar S^1 = S^1/B^{\gamma }$ is a $\P^{\gamma }$-local martingale.
\ehyp

In a typical market model (say, the Black-Scholes model), this assumption will
be satisfied, due to Girsanov's theorem, but it does not mean that the process $B^{\gamma }$ has any specific relationship to our model. When referring to results from \cite{PPB12}, we will sometimes interpret $\gamma $ as a virtual `risk-free rate', but this
interpretation is completely arbitrary and it does not have any bearing on the validity of results presented below.
We now define an auxiliary process $V^{\gamma } (\phi ) $ associated with an arbitrary self-financing trading strategy $\phi $.

\bd
Let $\phi $ be a self-financing trading strategy with the wealth process $V(\phi )$.
Then the process $V^{\gamma } (\phi ) $ is defined by the following formula
\be \lab{gamm1}
V^{\gamma }_t (\phi ) := V_t (\phi ) + B^{\gamma}_t \int_0^t (\gamma_u - r^0_u) \psi^0_u B^0_u (B^{\gamma }_u)^{-1} \, du .
\ee
\ed

Of course, the process $V^{\gamma }_t (\phi )$ does not represent the wealth of a self-financing strategy, in general.

\brem
The process $V^{\gamma } (\phi ) $ can be equivalently defined by
\be \lab{gamm1}
V^{\gamma }_t (\phi ) := V_t (\phi ) + B^{\gamma}_t \int_0^t (\gamma_u - r^0_u) (V_u (\phi )
 - \xi^1_u S^1_u) (B^{\gamma }_u)^{-1} \, du .
\ee
Unless $\gamma $ is interpreted as the risk-free rate, no financial interpretation of the last term in the formula above
is available.
\erem

Let us define the process $\bar V^{\gamma } (\phi )$ by setting $\bar V^{\gamma } (\phi ) = V^{\gamma } (\phi )/B^{\gamma }$.
 The following proposition shows that, for any self-financing trading strategy $\phi $, the process  $\bar V^{\gamma } (\phi )$ enjoys the martingale property under the probability measure $\P^{\gamma }$. This is a purely mathematical result and it does
not mean that $\P^{\gamma }$ is a `risk-neutral probability' in any sense, in general.

\bl \lab{gamprof}
Let $\phi $ be a self-financing trading strategy and let the process $V^{\gamma } (\phi )$ be given by \eqref{gamm1}.
Then the process $\bar V^{\gamma } (\phi )$ is a $\P^{\gamma }$-local martingale.
\el

\proof
For the sake of brevity, we drop $\phi $ from the notation $V(\phi )$ and $V^\gamma (\phi )$.
It suffices to show that
\bde
d\bar V^{\gamma }_t = \xi^1_t \, d \bar S^1_t
\ede
or, equivalently,
\be \lab{gamm2}
dV^{\gamma }_t  - \gamma_t V^{\gamma }_t \, dt = \xi^1_t \,  \big( dS^1_t - \gamma_t S^1_t \, dt \big).
\ee
By applying the It\^o formula to \eqref{gamm1}, we get
\begin{align*}
dV^{\gamma }_t &= dV_t + (\gamma_t - r^0_t)  \psi^0_t B^0_t \, dt +\frac{ V^{\gamma}_t - V_t }{B^{\gamma}_t } \, dB^{\gamma }_t  \\
&= dV_t + (\gamma_t - r^0_t) ( V_t - \xi^1_t S^1_t ) \, dt + \gamma_t ( V^{\gamma}_t - V_t ) \, dt .
\end{align*}
Therefore, using the self-financing property of $\phi $, we obtain
\begin{align*}
dV^{\gamma }_t  - \gamma_t V^{\gamma }_t \, dt  &= dV_t - r^0_t V_t \, dt - (\gamma_t - r^0_t) \xi^1_t S^1_t \, dt \\
&= dV_t - r^0_t (\xi^1_t S^1_t + \psi^0 B^0_t) \, dt - (\gamma_t - r^0_t) \xi^1_t S^1_t \, dt
\\ &=  dV_t - r^0_t \psi^0 B^0_t \, dt  - \gamma_t \xi^1_t S^1_t \, dt
\\ &=  \xi^1_t \, dS_t + \psi^0 \, dB^0_t  + r^0_t \psi^0 B^0_t \, dt   - \gamma_t \xi^1_t S^1_t \, dt
\\ &= \xi^1_t \, \big( dS^1_t - \gamma_t S^1_t \, dt \big)
\end{align*}
as was required to show. \endproof

Of course, when the equality $\gamma = r^0$ holds then Lemma  \ref{gamprof} reduces to Lemma \ref{oop}.  Lemma  \ref{gamprof} can
 thus be seen as an extension of  Lemma \ref{oop} to the general case when the discount factor is not necessarily a traded asset.
In the final step, we will illustrate Theorem 4.3 in \cite{PPB12}. Note, however, that a counterpart of
formula \eqref{gamm5a} is postulated in \cite{PPB12}, whereas we derive it from fundamentals.

\bhyp \lab{ahh3}
Assume that a contract has a single cash flow $X$ at time $T$ and a replicating self-financing strategy $\phi $ for $X$ exists.
\ehyp

Under suitable integrability assumption, the discounted wealth process
$\wt V(\phi )$ of a replicating strategy is a $\PS$-martingale. Consequently, the arbitrage price of $X$ can be computed using
the risk-neutral valuation formula, specifically,
\be \lab{gamm4}
\pi_t (X) = \EPS ( X  B^0_t (B^{0}_T)^{-1} \, | \, \F_t ).
\ee
Let us now take any process $B^{\gamma }$ such that the probability measure $\P^{\gamma }$ is well defined.
From Proposition  \ref{gamprof}, we deduce the following corollary showing that  $\P^{\gamma }$ can also be used
as a `pricing measure' after suitable adjustments of cash flows.

\bcor
If Assumptions \ref{ahh1}-\ref{ahh3} are satisfied then the price $\pi_t(X) = V_t(\phi )$ is also given by the following formula
\be  \lab{gamm5a}
\pi_t (X) = \EPG \Big( X B^{\gamma}_t (B^{\gamma }_T)^{-1}
+ B^{\gamma}_t \int_t^T (\gamma_u - r^0_u) \psi^0_u B^0_u (B^{\gamma }_u)^{-1} \, du  \, \Big| \, \F_t \Big).
\ee
Since
\bde
\psi^0_u B^0_u =  V_u (\phi ) - \xi^1_u S^1_u = \pi_u (X) - \xi^1_u S^1_u,
\ede
 formula \eqref{gamm5a} can also be rewritten as follows
\be \lab{gamm5}
\pi_t (X) = \EPG \Big( X B^{\gamma}_t (B^{\gamma }_T)^{-1}
+ B^{\gamma}_t \int_t^T (\gamma_u - r^0_u) (\pi_u (X) - \xi^1_u S^1_u) (B^{\gamma }_u)^{-1} \, du  \, \Big| \, \F_t \Big).
\ee
\ecor

\proof
It suffices to show that the right-hand side in  \eqref{gamm5a} coincides with $V(\phi )$ where a strategy $\phi $ replicates $X$.
The martingale property of $\bar V^{\gamma } (\phi )$ under $\P^{\gamma }$ means that, for all $t \in [0,T]$,
\be \lab{uu8}
\bar V^{\gamma }_t (\phi ) = \EPG \big( \bar V^{\gamma }_T (\phi ) \, | \, \F_t \big).
\ee
In view of \eqref{gamm1} and the equality $V_T(\phi )=X$, equality \eqref{uu8} implies that
\bde 
V_t (\phi )(B^{\gamma}_t)^{-1} + \int_0^t (\gamma_u - r^0_u) \psi^0_u B^0_u (B^{\gamma }_u)^{-1} \, du =
\EPG \Big( X (B^{\gamma}_T)^{-1} + \int_0^T (\gamma_u - r^0_u) \psi^0_u B^0_u (B^{\gamma }_u)^{-1} \, du \, \Big| \, \F_t \Big).
\ede
This immediately yields the asserted formula.
\endproof

As was already mentioned, a version of formula \eqref{gamm5} was postulated in  \cite{PPB12}  as a valid valuation recipe under funding costs (see the first formula in Section 4.5.1 in \cite{PPB12}). In our opinion, the arguments put forward in \cite{PPB12}, although they sometimes lead to a correct valuation result, are too complicated and they may require some guesswork for
finding the adjustment of cash flows. It is much simpler, as it appears, to start with a market model in which
 `instrumental variable' (e.g., a non-traded risk-free short-term rate) is not employed at all. Needless to say that formula \eqref{gamm4} is much easier to establish and implement than \eqref{gamm5}, so there is no practical advantage of using the latter representation for the numerical pricing purposes.


\end{document}